\documentclass[12pt,preprint]{aastex}

\def\mv{M_V}
\def\msun{$M_{\odot}$}

\def\lsun{$L_{\odot}$}

\def\hbeta{H$\beta$}
\def\hgamma{H$\gamma$}

\def\Te{T_{\rm eff}}
\def\logg{\log g}
\def\gta{\lower 0.5ex\hbox{$\buildrel > \over \sim\ $}} 
\def\lta{\lower 0.5ex\hbox{$\buildrel < \over \sim\ $}} 

\slugcomment {Submitted to ApJSuppl } 

\begin{document}

\title{THE FORMATION RATE, MASS AND LUMINOSITY FUNCTIONS 
OF DA WHITE DWARFS FROM THE PALOMAR GREEN SURVEY}

\author{James Liebert }
\affil{Steward Observatory, University of Arizona, Tucson AZ 85726} 
\email{liebert@as.arizona.edu}  

\author{P. Bergeron }
\affil{D\'epartement de Physique, Universit\'e de Montr\'eal,}
\affil{C.P. 6128, Succ. Centre-Ville, Montr\'eal, Qu\'ebec,} 
\affil{Canada H3C 3J7 }
\email{bergeron@astro.umontreal.ca}

\author{and}

\author{J.B. Holberg} 
\affil{Lunar and Planetary Laboratory, University of Arizona}
\affil{Tucson, AZ 85726} 
\email{holberg@argus.lpl.arizona.edu}

\begin{abstract}

Spectrophotometric observations at high signal-to-noise ratio were
obtained of a complete sample of 347 DA white dwarfs from the Palomar
Green (PG) Survey.  Fits of observed Balmer lines to synthetic spectra
calculated from pure-hydrogen model atmospheres were used to obtain
robust values of $\Te$, $\logg$, masses, radii, and cooling ages.  The
luminosity function of the sample, weighted by $1/V_{\rm max}$, was
obtained and compared with other determinations.  Incompleteness of the
sample due to selection by photographic $m_u$--$m_b$ color and magnitude
limits was found to be a serious problem, and an attempt is made to
correct for this.  The mass distribution of the white dwarfs is derived,
after important corrections for the radii of the white dwarfs in this
magnitude-limited survey and for the cooling time scales.  This
distribution has (1) a ``peak'' component centered near 0.6~\msun, (2) a
low mass component centered near 0.4~\msun, and (3) a high mass
component above about 0.8~\msun.  The formation rate of DA white dwarfs
from the PG is estimated to be $0.6\times10^{-12}\ {\rm pc}^{-3}\ {\rm
yr}^{-1}$.  Of these, 75\% are from the peak component, 10\% from the
low mass component, and 15\% from the high mass component.  The low mass
component requires binary evolution for 100\% of the objects, with a
degenerate companion likely in the majority of cases.  Comparison with
predictions from a theoretical study of the white dwarf formation rate
for single stars indicates that $\ge$80\% of the high mass component
requires a different origin, presumably mergers of lower mass double
degenerate stars.  The need for a binary channel may not be as great for
the massive, very hot white dwarfs found in the EUV all-sky surveys.  In
an Appendix, we even suggest that an enhanced density of the massive
white dwarfs at lower Galactic latitudes might be due to some of them
being the progeny of B stars in Gould's Belt.  In order to estimate the
recent formation rate of all white dwarfs in the local Galactic disk,
corrections for incompleteness of the PG, addition of the DB-DO white
dwarfs, and allowance for stars hidden by luminous binary companions had
to be applied to enhance the rate.  An overall formation rate of white
dwarfs recently in the local Galactic disk of $1\pm0.25\times10^{-12}\
{\rm pc}^{-3}\ {\rm yr}^{-1}$ is obtained.  Admittedly, the systematic
errors in this estimate are difficult to quantify.  Two recent studies
of samples of nearby Galactic planetary nebulae lead to estimates around
twice as high.  Difficulties in reconciling these determinations are
discussed.

\end{abstract}


\section{INTRODUCTION}

The luminosity and mass functions (LF/MF) derived from a complete sample
of hot white dwarfs in the solar neighborhood contain a variety of
useful information on the star formation history of the local Galactic
disk and on stellar evolution.  First, the LF is a direct measure of the
current death rate of stars in the local disk.  Secondly, the mass
distribution of a volume-defined and complete sample reveals the amount
of mass lost during stellar evolution from an initial stellar mass
distribution.  Moreover, the MF can also reveal the roles of close
binary evolution in forming some of the white dwarfs.

The Palomar Green (PG) Survey \citep{PG} provides a complete, magnitude
and photographic $m_u$--$m_b$ color limited survey covering nearly one
quarter of the sky at Galactic latitudes $b \geq$ 30 degrees. The
estimation of the DA white dwarf formation rate and luminosity function
from the entire area of this survey \citep{fleming86} utilized
photographic, image tube and vidicon spectra, Str\"omgren and
multichannel spectrophotometry, and correspondingly limited analysis
techniques -- i.e. the assumption of a mean mass of 0.6~\msun, and the
estimates of $\Te$ using a variety of colors and \hbeta\ equivalent
width measurements.  Moreover, a significant fraction of stars in the PG
Survey classified originally as DA white dwarfs turn out to be lower
surface gravity stars (and a few migrated the other direction).

In this reanalysis, all valid DA stars from the survey were reobserved
with optical CCD spectra.  Hydrogen Balmer line profiles through \hbeta\
were fit with model atmosphere models to determine the effective
temperature and surface gravity of the star and estimate the mass
following the procedures of \citet[][hereafter BSL]{bsl}.  The
evolutionary models of \citet{wood90,wood95} were used to provide a
second relation between radius and mass at the derived $\Te$, so that
individual stellar parameters for each star could be obtained.  For
several stars in this sample not reobserved, we will utilize similar
fits by other research groups (see below), with which we will otherwise
compare our results.

Various other, large samples of hot DA stars have recently been
analyzed using primarily the Balmer line fitting technique to obtain
the stellar and atmospheric parameters.  \citet{finley97} analyzed 174
stars selected from catalogs of known white dwarfs and those
newly-discovered in space-based extreme ultraviolet radiation surveys.
Similarly, \citet{marsh97} performed a similar analysis of
the optical spectrophotometry of the sample of 89 hot DA stars
discovered in the ROSAT all-sky survey, although many of these
detections were previously known stars.  Likewise, \citet{vennes97b}
analyzed 110 stars discovered or rediscovered in the
similar all-sky survey of the {\it Extreme Ultraviolet Explorer} (EUVE),
with special emphasis on analyzing the mass distribution, and making
comparison with stars selected optically (such as this study).  We
shall compare our parameter determinations with those from
the above studies, in order to determine if there are systematic
differences in the $\Te$ and $\logg$ values, along with the resulting
derived parameters.

The EUV-derived studies have demonstrated the wide variation in the EUV
opacity at a given $\Te$ among those with nearly-pure hydrogen
atmospheres, vs. those with strong traces of heavier elements.  This of
course has called into question the ability to use an EUV-selected
sample to define a volume-limited, complete sample of white dwarfs --
cf. \citet{finley97,marsh97}.  However, we will bring possible new
insight on this question in this paper.

In this paper, we shall review in \S~2 the input data set, model
atmosphere assumptions and physics, and fitting technique.  In \S~2.4
the derived $\Te$, $\logg$, mass and radius values are presented, as are
the estimated cooling ages from the models of \citet{wood95} and
\citet{althaus01}.  Comparison of the overlapping samples with the
results of other studies is made in \S~2.5.  The LF and MF are presented
in \S~3.  It is useful to discuss the LF and other distribution
functions in terms of three mass components -- a low mass component
centered near 0.4~\msun, the dominant 0.6~\msun\ ``peak'' component, and
a high mass component at $>0.8$~\msun.  The completeness of the PG
sample, a difficult issue, is discussed in \S~4.  Then in \S~5 we arrive
at the most fundamental results of this analysis, the formation rates of
PG DA white dwarf stars for recent times in the local Galactic disk.
These rates are determined for the three mass components defined in
\S~3.  To estimate the total, recent formation rate of white dwarf stars
-- or the ``death rate'' of stars of low and intermediate masses -- it
is necessary to increment the results for DB/DO stars (\S~5.2), for
survey incompleteness (\S~5.3), and for white dwarfs hidden by luminous
companions (\S~5.4).  The total white dwarf formation rate is given in
(\S~5.5).  We then compare in \S~5.6 our formation rate estimate with
other determinations for white dwarfs and for planetary nebulae.  In
\S~6 we attempt a quantitative assessment of the fractions of the
component of massive white dwarfs formed from (1) single progenitors of
high initial mass, and (2) presumably, a binary white dwarf channel.  In
\S~7 we comment on objects of special spectral characteristics --
magnetic white dwarfs, DAB stars in and near 30-45,000~K (the so-called
``DB gap'') and pulsating ZZ~Ceti stars.  Finally, in \S~8 is a brief
summary of this study, and its likely value as a ``benchmark'' for
future studies of much larger samples of hot white dwarfs.

\section{OBSERVATIONS, MODELS, FITTING TECHNIQUE, AND DERIVED PARAMETERS}

\subsection{Sample Selection} 

Our goal was to observe with current CCD detectors at high S/N ratio, a
complete sample of DA stars from the PG Survey. The PG catalog contains
a total of 420 DA stars, 363 of which are part of the complete
sample\footnote{For the PG DA stars discussed in this paper, finding
charts, accurate coordinates, and summaries of prior spectroscopic
analyses are available on the White Dwarf Database at
http://procyon.lpl.arizona.edu/WD/}.  Upon reobservation, 27 of these
were in fact misclassifications, generally turning out to be subdwarf B
or OB stars including at least one with a G-K companion (PG~0009$+$191,
0026$+$136, 0111$+$177, 0221$+$217, 0910$+$622, 0947$+$639, 1025$+$258,
1035$+$532, 1121$+$145, 1134$+$124, 1236$+$479, 1247$+$554, 1302$+$284,
1308$-$099, 1347$+$254, 1430$+$427, 1450$+$432, 1525$+$422, 1538$+$269,
1539$+$530, 1630$+$249, 2204$+$071, 2246$+$154, and 2333$-$002),
subdwarf O stars (PG 0035$+$125 and 1138$+$424), and a DC white dwarf
(PG 1055$-$073).  One of these, PG~2204$+$071, has two entries in the PG
catalog with almost identical coordinates, classified DA and DA5; only
the southern and fainter object is a genuine DA star while the other
star corresponds to a sdOB star.  Major studies of sdB \citep{saffer94}
and sdO stars \citep{thejll94}, and the identification of candidate B
main sequence stars at high galactic latitude \citep{saffer97} led to
the reobservation of most PG stars originally classified as lower
gravity objects. Twelve of these proved to be DA white dwarfs (and one a
DO white dwarf). Thus, while we cannot guarantee that a few DA stars do
not continue to lurk among the samples originally classified sdB, sdO or
otherwise, we are convinced that the number is a few at most.  A more
important issue, to be addressed later, is whether the PG catalog is
itself significantly incomplete.

We ended up with a complete sample of 347 genuine DA stars to be
analyzed using the spectroscopic technique. Optical spectra for this
sample have been gathered over the last 10 years from various
sources. About 20\% of them have been kindly provided to us by C.~Moran
(private communication), while the rest has been secured using the 2.3 m
telescope at the Steward Observatory Kitt Peak Station, equipped with
the Boller \& Chivens spectrograph and a Texas Instrument CCD detector
with 15$\mu$ pixels. The spectral coverage is about
$\lambda\lambda$3100--5300, thus covering \hbeta\ up to H9 at an
intermediate resolution of $\sim6$~\AA\ FWHM. The spectra from C.~Moran
have a similar spectral coverage but a higher resolution of 3~\AA\
FWHM. Over 100 stars have been observed twice or more in order to
estimate the external error of the atmospheric parameter determinations
(see below). No spectroscopic observations were obtained for the
strongly magnetic white dwarfs PG~1031$+$234 \citep{schmidt86} and
PG~1533$-$057 \citep{liebert85} since these cannot be analyzed within
our theoretical framework, and the atmospheric parameters for these
stars will thus be taken from the literature.

The distribution of signal-to-noise (S/N) ratios for our 345 optical
spectra is displayed in Figure \ref{fg:sn}.  The S/N is based on
estimates of the RMS noise per pixel in the continuum.  About 50\% of
our sample has ${\rm S/N}>80$ and 90\% above 50. Only the faintest
objects in the sample have a lower signal-to-noise ratio. Sample spectra
covering the entire temperature range of the PG sample are displayed in
Figure \ref{fg:sample}. Effective temperatures range from above
100,000~K down to $\sim 7000$ K. Examples of DAO (PG~1252$+$378),
magnetic (PG~2329$+$267), and composite (PG~1314$+$294) white dwarfs are
shown.  Note the strong contrast between the high gravity white dwarf
PG~1058$-$129 and the adjacent normal gravity white dwarfs of comparable
effective temperatures, PG~1553$+$354 and PG~1713$+$333.

\subsection{Model Spectra} 

The sensitivity of Balmer line profiles to $\Te$ and $\logg$ has been
known for a long time \citep[c.f.][]{wegner81}.  This was first
exploited using new generations of both CCD spectrophotometry and
models incorporating the \citet{HM88} occupation probability formalism
for hydrogen by \citet{bergeron90,bergeron91}. Refinements to the
Hummer \& Mihalas formalism were utilized in the first major study of
warm/hot DA white dwarfs by BSL. The BSL study of the mass 
distribution of white dwarfs was generally restricted to the
$\Te$ range 15,000--40,000~K because of problems with the accuracy of
the models both above and below this range.

For the cooler stars, the problem is posed by the onset of significant
atmospheric convection.  The derived $\Te$ and $\logg$ values are
sensitive to the input parameters of the mixing length theory, though
\citet{bergeron95} find that the ML2/$\alpha=0.6$ formulation appears 
to yield the best internal consistency between optical and UV
effective temperatures, trigonometric parallaxes, $V$ magnitudes, and
gravitational redshifts. This formulation is therefore adopted for
this study, which includes stars as low in temperature as
approximately 7000~K.

For stars hotter than about 40,000~K several complications must be
acknowledged that compromise the accuracy of these results, which
generally assume pure hydrogen atmospheres unless helium is visible in
the spectrum (a DAO, or DAB classification).  First, even if helium is
not detected in the form of a He~II 4686~\AA\ line, it may be present
in an abundance sufficient to affect significantly the atmospheric
structure and Balmer line profiles of a very hot DA star according to
the LTE analysis of \citet{bergeron94}. However, \citet{napi97} has
shown that this conclusion is an artifact caused by the assumption of
LTE, and the influence of small traces of helium on the Balmer
profiles vanishes when NLTE effects are taken into account. Napiwotzki
thus recommends using pure hydrogen models if an LTE analysis of the
Balmer lines is to be performed, which is what we are assuming here.

A more complex issue is posed by the discovery of trace abundances of
elements heavier than helium -- iron and other heavy metals, CNO, and Si
in particular -- in the atmospheres of most DA stars generally hotter
than 50,000~K, in particular in the DAO white dwarfs.  High dispersion
spectra with the {\it International Ultraviolet Explorer} resulted in
the detections of various ions in DA stars.  It is impossible here to
cite the large number of papers responsible for this discovery
\citep[see][]{hbs98}.  However, the discovery of unexpectedly-large
opacities at the extreme ultraviolet (EUV) wavelengths measured with
{\it ROSAT} and {\it EUVE} played a pivotal role
\citep[cf.][]{marsh97,vennes97b}.  At the same time, \citet{napi92}
encountered a problem in fitting simultaneously the Balmer lines of hot,
hydrogen-rich DA, DAO and H-rich planetary nebulae central stars; it now
appears that elements heavier than helium need to be accounted for, in
order to get correct fits and stellar parameters
\citep{bergeron93,werner96}.  The sense of the problem is that the $\Te$
value obtained by fitting the synthetic line profiles of a pure-H
atmosphere is higher than the $\Te$ obtained with additional heavy
element opacities \citep[]{barstow03b}, at least for NLTE models.  The
problem is obviously complex, since a large number of
radiatively-levitated ions might be contributing, but there is
insufficient spectrophotometric information to identify these
individually.  The dependence of $\logg$ on the heavy element opacities
may be more complicated.

It should be noted there presently also exists a significant discrepancy
between $\Te$ results obtained from the type of Balmer line analyses
used in this paper and analogous methods used to fit the Lyman line
profiles in very hot DA white dwarfs.  \citet{barstow03a} have analyzed
the Lyman line profiles of 16 DA white dwarfs observed with the {\it Far
Ultraviolet Spectroscopic Explorer} ({\it FUSE)} and compared these
results with temperature determined from the Balmer lines of the same
stars.  Specifically, it was found that marked systematic differences
between the two methods occur for those stars with $\Te>50,000$~K.
The sense of this difference is that the Lyman $\Te$ values are
$\sim10$\% higher than the corresponding Balmer values.  At lower
temperatures, the two methods are in very good accord.

The DA white dwarfs from the complete PG sample cover a wide range of
effective temperatures including hot stars where NLTE effects on Balmer
line profiles are important \citep{napi97}, and cooler stars where
energy transport by convection dominates. Even though NLTE model
atmosphere codes are now becoming widely available \citep[e.g., TLUSTY
and SYNSPEC;][]{hubeny95}, these codes are usually not able to handle
convection satisfactorily. Our strategy was thus to adopt our LTE model
atmosphere code (see BSL and references therein), which can handle
convection reliably, up to an effective temperature where NLTE effects are
still negligible and the atmospheres are completely radiative, and then
switch to the TLUSTY and SYNSPEC packages to deal with NLTE effects at
higher temperatures.

First, we need to ensure that at the branching point, both model
atmosphere codes yield similar atmospheric structures and model
spectra. I. Hubeny (private communication) kindly calculated for us LTE
and NLTE pure hydrogen models and spectra for $\Te>20,000$~K using
TLUSTY and SYNSPEC. In an experiment similar to that shown in Figure 3
of \citet{napi99}, we fitted the Balmer lines of the LTE models with the
NLTE spectra to derive the LTE corrections displayed in Figure
\ref{fg:NLTEcorr}; these represent the corrections that would need to be
applied to the atmospheric parameters obtained under the assumption of
LTE models. Our results are qualitatively in agreement with the results
of Napiwotzki et al. (Fig.~3), although the effects in $\logg$ found at
high effective temperatures are significantly smaller here (note that in
Figure 3 of Napiwotzki et al.~the offsets are magnified 3 times in {\it
both} $\Te$ and $\logg$). Thus, LTE models tend to overestimate both
$\Te$ and $\logg$, although the effects in $\logg$ are practically
negligible ($\lta 0.03$ dex).  As illustrated in {\it our} Figure
\ref{fg:NLTEcorr}, the onset of differences due to NLTE becomes
significant about 40,000~K.

The next step was to fit our own grid of LTE model spectra with those
calculated with TLUSTY and SYNSPEC under the assumption of LTE.
Surprisingly, the differences in $\Te$ and $\logg$ found initially were
much larger than the LTE/NLTE differences observed in Figure
\ref{fg:NLTEcorr}! The origin of these differences was traced back to
the use of different tables of Stark broadening for the hydrogen lines
-- our LTE models now make use of the extended calculations of
\citet{lemke97}, as does SYNSPEC -- to the details of the implementation
of the Hummer \& Mihalas occupation probability formalism
\citep{bergeron93b}, and more importantly, to the number of depth points
and frequency points used in the model calculations. So even though NLTE
effects are indeed important for the analysis of hotter DA stars (see
Fig.~\ref{fg:NLTEcorr}), attention to details such as those listed above
may be equally important.

After the differences between the LTE/TLUSTY code and our own LTE code
were understood and resolved, the LTE synthetic spectra obtained from
both codes agreed to better than 1\% in $\Te$, and 0.02 dex in $\logg$,
from $\Te=20,000$~K to 90,000~K. Thus, the effective temperature at
which the two model grids were matched was set at 20,000~K where the
convective flux is zero, and where the LTE approximation holds.  The
NLTE switch in TLUSTY was then turned on to calculate model spectra
above $\Te=20,000$~K, while our own LTE code was used to calculate
cooler (convective) models. Our complete pure hydrogen model grid covers
a range between $\Te=1500$~K and 140,000~K by steps of 500 K at low
temperatures ($\Te<17,000$~K) and 5000 K at high temperatures
($\Te>20,000$~K), and a range in $\logg$ between 6.5 and 9.5 by steps of
0.5 dex (steps of 0.25 dex were used between 8000 K and 17,000 K where
Balmer lines reach their maxima). We thus end up with a homogeneous
model grid that includes consistently NLTE effects as well as convective
energy transport, and completely covers the range of parameters in the
PG sample.

\subsection{Fitting Technique} 

Our fitting technique is an improved version of that used by BSL. The
first step is to normalize the flux from an individual line, in both
observed and model spectra, to a continuum set to unity at a fixed
distance from the line center. The comparison with model spectra, which
are convolved with the appropriate Gaussian instrumental profile (3, 6,
or 9~\AA), is then carried out in terms of these line shapes only. The
most sensitive aspect of this fitting technique is to define the
continuum of the observed spectra. Here we rely on the procedure
outlined in \citet{bergeron95} where the observed spectrum is fitted
with several pseudo-Gaussian profiles \citep[see also][]{saffer88} using
the nonlinear least-squares method of Levenberg-Marquardt
\citep{press86}. The normal points are then fixed at the points defined
by this smooth function. As discussed by \citet{bergeron95}, this method
is much more accurate when a glitch is present in the spectrum at the
location where the continuum is set. It also provides a precise value of
the line center which can be corrected to the laboratory wavelength.

This procedure is quite reliable in the temperature range where the
Balmer lines are strong and can be approximated as a sum of
pseudo-Gaussian profiles ($16,000\ \gta\Te\ \gta9000$~K). Outside this
temperature range the method becomes more unstable when the continuum
between the lower Balmer lines becomes linear, or when the higher Balmer
lines vanish. There, instead of using pseudo-Gaussian profiles, we rely
on theoretical spectra to reproduce the observed spectrum, including a
wavelength shift, a zero point, as well as several order terms in
$\lambda$ (up to $\lambda^6$). The normal points are then fixed at the
points defined by this smooth model fit.  Note that the values of
$\logg$ and $\Te$ at this stage are meaningless since too many
fitting parameters are used, and the model just serves as a smooth
fitting function to define the continuum of the observed
spectrum. Examples of continuum fitting using the two procedures
described above are shown in Figure \ref{fg:f4}.

Once the Balmer lines are normalized to a continuum set to unity, we use
our grid of model spectra to determine $\Te$ and $\logg$ in terms of
these normalized profiles only. Our minimization technique again relies
on the nonlinear least-squares method of Levenberg-Marquardt, which is
based on a steepest descent method. We consider here only \hbeta\ to H8
in the fitting procedure. For cases where the red portion of the
spectrum is contaminated by an unresolved companion, we neglect \hbeta\
in the fitting procedure (PG~0014$+$098, 0805$+$655, 0933$+$026,
1026$+$002, 1037$+$512, 1049$+$103, 1314$+$294,
1443$+$337, 1622$+$324, 1646$+$062), and in some cases even
\hgamma\ (0004$+$061, 0824$+$289, 0950$+$186, 1001$+$204,
1210$+$464, and 1643$+$144).

As is well known, when the effective temperature of the white dwarf is
close to the region where the equivalent widths of the Balmer lines
reach a maximum \citep[$\Te\sim 13,500$~K; see Fig.~4 of][]{bergeron95},
two solutions are possible, one on each side of the maximum. Even though
the optical spectra are not spectrophotometric in the sense that their
slopes are accurate enough to serve as a precise temperature indicator,
they can still be used to discriminate between the cool and the hot
solutions. Hence for such stars in our sample, we examine the two
solutions obtained by using a cold and a hot seed, compare the model
spectrum with the observed spectrum, both normalized at 4600~\AA, and
pick the solution that best reproduces the observed energy distribution.

Effective temperatures and surface gravities are obtained in this manner
for all 347 DA stars in the PG sample. A small number of objects in our
sample have been analyzed in greater detail in the literature than
within our current framework, in which cases we substitute our $\Te$ and
$\logg$ values with those taken from these studies. This is the case for
the magnetic white dwarfs PG~1031$+$234 \citep{schmidt86}, 1220$+$234
\citep{liebert03}, 1533$-$057 \citep{liebert85}, and 1658$+$441
\citep{schmidt92}, the DAO stars 0134$+$181, 0823$+$317, 0834$+$501,
0846$+$249, 1202$+$608, 1210$+$533, 1214$+$268, 1252$+$378, and
1305$-$017 \citep{bergeron94}, the unresolved double degenerate systems
PG~0945$+$246 \citep[LB~11146,][]{liebert93} and 1115$+$166
\citep{bergeron02}.

\subsection{Results}

The atmospheric parameter determinations for the 347 DA stars from the
complete PG sample are reported in Table~1; the values in parentheses
represent the internal errors of the fitting technique for $\Te$ and
$\logg$. These represent only the formal uncertainties of the fitted
atmospheric parameters obtained from the covariance matrix (see BSL and
Press et al.~1986 for details). Sample fits covering the temperature
range of our sample are displayed in Figure \ref{fg:f5}. In
addition, we provide in Table 2 the atmospheric parameters of several DA
stars in the PG survey that were also observed but which are not part of the
complete sample.

Other quantities in Table~1, derived from the evolutionary white dwarf
models, include the stellar mass ($M$) and uncertainty, the absolute
visual magnitude ($\mv$), the luminosity ($L$), the $V$ magnitude, the
distance ($D_{\rm pc}$), the $1/V_{\rm max}$ weighting (pc$^{-3}$), and the
log of the cooling time ($\log\tau$).

The $V$ magnitudes are important in estimating the distances of the
stars.  There exists published non-photographic photometry for 291 of
the PG DA stars.  For these we have determined a mean $V$ band magnitude
using broadband $V$ or equivalent Str\"omgren $y$ magnitudes from
\citet{mccook99}.  In some cases obviously discrepant magnitudes were
excluded.  In general Palomar multi-channel spectrophotometry
\citep[]{oke69} was not used, unless it was the only available source of
photometry.  In those cases the multi-channel $v$ \citep[]{jlg76} was
converted to $V$ using the relation $V$ = $v$ + 0.07.  For 12 PG stars
we were able to use DR~1 or DR~2 Sloan Digital Sky Survey (SDSS)
photometry to determine an equivalent $V$ from the transformation of the
SDSS $g$ and $r$ band magnitudes taken from \citet{smith02}; these are
noted in Table 2.  For 43 stars only the original PG photographic $m_b$
magnitudes were available.  For these stars we estimated an effective
$V$ magnitude by using the larger sample of PG stars to form a
correlation between the spectroscopic $\mv$ and the pseudo color
$m_b$--$V$.  As expected there is a substantial dispersion in the $m_b$
values (standard deviation, 0.477 magnitudes) and consequently our $V$
band estimates for these stars are quite uncertain.  Values for these
stars are given only to tenths of the magnitude, and they are also noted
in Table 2.

Since the calculations of cooling times require the use of evolutionary
models for white dwarfs, it is necessary to make an assumption about the
thickness of the outer hydrogen layer mass. While the weight of evidence
now is that most DA white dwarfs have ``thick'' outer hydrogen layers --
i.e., ``evolutionary'' layer masses of the order $q({\rm H})\equiv
M_{\rm H}/M_{\star}=10^{-4}$ \msun -- there is still reason to believe
that a minority of DA stars, in particular those becoming DB stars
below the 30,000--45,000~K DB gap, have much thinner outer
layers. Layer masses one or more orders of magnitude less than the
evolutionary value are better modeled using the assumption of zero
hydrogen layer mass, as was assumed in the BSL study. It is well known
that the assumption of an evolutionary H layer results in the
determination of a larger mass, and the effect is greater for the hotter
and/or lower mass stars.

Since the carbon-core models of \citet{wood95} with thick hydrogen
layers of $q({\rm H})=10^{-4}$ extend to much higher effective
temperatures than those of \citet{wood90} with no hydrogen layers, we
use the former models throughout. Low mass white dwarfs, however, are
likely to be composed of helium cores, for which we rely instead on the
evolutionary helium-core models of \citet{althaus01}. To illustrate the
parameter space used for the interpolation and to show also how these
different evolutionary models match in terms of mass, we plot in Figure
\ref{fg:f6} the $\Te$ and $\logg$ values for the complete PG
sample together with the evolutionary models of \citet{wood95} and
\citet{althaus01}. Unfortunately, the Althaus et al.~models go up to
only 0.406 \msun. Since there appears to be a paucity of stars near 0.45
\msun\ (see Fig.~\ref{fg:f6} and also our mass distribution
below), we fix arbitrarily our mass cutoff between carbon- and
helium-core models at 0.46 \msun. This ensures that the inferred masses
for the helium-core white dwarfs are still reasonably well extrapolated,
while remaining below the masses obtained for the white dwarfs with
slightly larger values of $\logg$ but inferred from the carbon-core
models of Wood.

The NLTE effects on the determination of white dwarf atmospheric
parameters can be estimated by analyzing all stars in our sample with
LTE models. The comparison between LTE and NLTE results is displayed
in Figure \ref{fg:f7}. The results show that NLTE effects are
always small for the analysis of DA stars, in agreement with the
conclusions reached by \citet{napi97}. The general trend is that LTE
solutions tend to overestimate both the effective temperatures and the
surface gravities. In some instances, the differences in $\logg$ are
significantly larger than those inferred from Figure \ref{fg:NLTEcorr}
using models only. These correspond to the white dwarfs that show the
so-called ``Balmer line problem'' first discussed by \citet{napi92},
for which neither the LTE nor the NLTE pure hydrogen models provide a
satisfactory fit to the observed hydrogen line profiles.

As discussed many places in the literature (see, e.g., BSL), the {\it
internal} errors of our fitting technique given in Table~1 can be made
arbitrarily small given that the signal-to-noise ratio is high, and that
the model spectra reproduce the observed data in detail. The true error
budget, however, is dominated by the {\it external} uncertainties
originating from the flux calibration in particular.  During the course
of this project, we have obtained multiple observations for many stars
in order to increase the overall signal-to-noise ratio of the sample,
and to make the spectroscopic data set as homogeneous as possible. As
such, we ended up with multiple (two spectra or more) observations for
126 white dwarfs in our sample, for a total of 284 spectra. Of course
for each star, we always pick the best spectrum in the final
analysis. Since all spectra have been obtained independently by
different observers, with an independent reduction, etc., these spectra
can be used to get an estimate of the external error of the atmospheric
parameter determinations. We thus fitted all 284 spectra using our
fitting technique, and calculated the average parameters and standard
deviations for each star, which are displayed in Figure
\ref{fg:multiple} as a function of $\Te$. The standard deviations in
$\Te$ are divided by the average temperature for each star to obtain a
distribution that is largely independent of $\Te$. We also show in Figure
\ref{fg:multiple} the {\it average} standard deviations of each
atmospheric parameter, 0.038 dex in $\logg$ and 1.2\% in $\Te$, which we
adopt as a measure of the external error of our fitting procedure.  Note
that these are even conservative estimates since each time, our best
spectrum is compared with admittedly worse spectra. Further note that we 
do not take into account possible systematic errors related to the model
atmospheres themselves.

\subsection{Comparison with Other Investigations}

As discussed in the introduction, Balmer line analyses of other large
samples of DA stars have been conducted by various investigators.  In
general these analyses have relied on Balmer line fitting techniques
similar to those described in this paper but have employed independent
observational data and independently computed model grids.  It is
therefore important to investigate the degree of consistency between our
$\Te$ and $\logg$ results and these unrelated data sets.  Such
comparisons also allow an assessment of the role of external systematic
errors between the various data sets, models and fitting techniques.

In Figure \ref{fg:f9} we compare our $\Te$ and $\logg$
determinations for PG stars which we have in common with the data sets
of \citet{finley97,vennes97b,marsh97,homeier98} and \citet{koester01},
respectively.  In each case we plot the quantities $T_{\rm
others}-T_{\rm PG}$ and $\logg_{\rm others}-\logg_{\rm PG}$ vs $T_{\rm
PG}$, where the subscripts refer to the above independent data sets and
our PG data set, respectively.  In order to estimate any systematic
offsets in temperature and gravity between our results and others, we
have calculated unweighted averages of the data in Figure
\ref{fg:f9}.  Unweighted averages were used because it is clear
in each case that both the uncertainties in both the temperature and
gravity are large with respect to the true level of mutual uncertainty
for individual stars. 

\subsubsection{Finley et al. Data}

\citet{finley97} determined the effective temperatures and surface
gravities of 174 DA stars having $\Te$ greater than 25,000~K.  These
stars were selected from a variety of catalog sources including DA stars
detected in the ROSAT and EUVE all-sky EUV surveys. The analysis of the
temperatures and gravities is based on pure hydrogen LTE atmospheres
derived from the grids of Koester.  The masses are calculated from the
thick models of \citet{wood95}.  There are 62 stars in common between
our PG sample and those of \citet{finley97}.  In Figure
\ref{fg:f9} we show the comparison of the Finley
et al.  $\Te$ and $\logg$ determinations with respect to our
results. The Finley et al. data are, on average, 1.7\% lower in
temperature and 0.06 dex lower in gravity than our results.

\subsubsection{Vennes et al. Data}

\citet{vennes99} describes the results for a purely EUV-selected sample
of 141 DA stars. The full $\Te$ and $\logg$ results for this sample are
variously contained in \citet{vennes96,vennes97a,vennes97b}, and
\citet{vennes99}.  Due to the EUV-selected nature of this sample the
$\Te$ values of all stars are effectively in excess of 25,000~K.  The
analysis of the temperatures and gravities is based on pure hydrogen LTE
atmospheres derived from the grids of \citet{vennes92}.  Masses are
calculated for both the thick and thin models of \citet{wood95}. There
are 17 stars in common between our PG sample and those of
\citet{vennes97b}.  In Figure \ref{fg:f9} we show
the comparison of the Vennes $\Te$ and $\logg$ determinations with
respect to our results. The Vennes et al.  data are on average 0.6\%
lower in temperature and 0.03 dex higher in gravity than our results.

\subsubsection{Marsh et al. Data}

\citet{marsh97} describe the results for a purely EUV-selected sample of
89 DA stars.  As with the Vennes sample, all stars are effectively in
excess of 25,000~K. The analysis of the temperatures and gravities is
based on pure hydrogen LTE atmospheres derived from the grids of
Koester, while the masses are calculated from the thick models of
\citet{wood95}.  There are 16 stars in common between our PG sample and
those of \citet{marsh97}.  In Figure \ref{fg:f9}
we show the comparison of the Marsh et al. $\Te$ and $\logg$
determinations with respect to our results. The Marsh et al. data, which
show the most variance of all data sets, are on average 3.3\% lower in
temperature and 0.10 dex lower in gravity than our results.

\subsubsection{Homeier et al. Data} 

\citet{homeier98} describe the results for an initial sample of 80 DA
stars drawn from the Hamburg Quasar Survey, an objective prism survey
of at high northern galactic latitudes.  The Homeier et al.~sample
covers the temperature range above 10,000~K. The analysis of the
temperatures and gravities is based on pure hydrogen LTE atmospheres
derived from the grids of Koester while the masses are calculated from
the thick hydrogen models of \citet{wood95}. There are 9 stars in
common between our PG sample and those of Homeier et al.  In Figure
\ref{fg:f9} we show the comparison of the
Homeier $\Te$ and $\logg$ determinations with respect to our
results. The Homeier et al. data are on average 0.3\% higher in
temperature and 0.08 dex lower in gravity than our results.

\subsubsection{Koester et al. Data} 

The SPY project is a large, general observational program to find close
binaries from several white dwarf surveys, mainly as potential
progenitors for SNIa \citep[]{napi01}. \citet{koester01} analyze stars
from the SPY project in a temperature range between about 8,000 and
30,000~K.  There are 17 stars in common with our sample.  In Figure
\ref{fg:f9} we show the same comparisons with respect to our results.
The Koester et al. data are on average 0.6\% higher in temperature and
0.08 dex lower in gravity than our results.

\section{LUMINOSITY, MASS AND OTHER DISTRIBUTION FUNCTIONS}

In \citet{liebert95} and especially in \citet{blr01}, the importance of
combining the mass and luminosity functions (MFs, LFs) into one
distribution function for the comparison of cooling times is emphasized.
The primary parameter we will be seeking is the formation rate of white
dwarfs in the recent history of the local Galactic disk.  However, for
comparison with the traditional, published LFs, we present first the
derived LF for all 347 stars.

\subsection{A Comparison of Luminosity Functions} 

The LF was calculated using the method of \citet{green80} and
\citet{fleming86}.  Briefly, given the limiting magnitude of the
particular field in which a given object was found, we let $V_{\rm
max}$ be the volume defined by the maximum distance at which a given
object would still appear in the sample.  To correct for the
nonuniform distribution of stars (i.e. the Galactic disk scale height),
we define $dV^z=\exp(-z/z_0)\,dV$, where $z_0$ is the assumed scale
height.  Then each star's contribution to the local space density is
$1/V_{\rm max}$.

In Figure \ref{fg:f10} the visual luminosity function of all DA stars
in the complete PG sample is presented in half-magnitude bins, assuming
a scale height for the Galaxy of $z_0=250$ pc \citep[as
did][]{fleming86}. The dotted line represents the results of
\citet{fleming86}, and the dashed line that for the KUV sample of
\citet{darling94} to be discussed later.  Note that the color correction
factors given in Table 2 of Fleming et al.~are not included in our
calculations. To within the errors, which are of order 30\% and larger
in a $1/V_{\rm max}$ calculation, the two earlier and present PG LFs are
indistinguishable.  The total number of DA white dwarfs per 1000
pc$^{-3}$ is again $0.50\pm0.05$ for $\mv<12.75$.  The close agreement
of the two determinations might seem surprising, given (1) the
appreciable number of misclassified white dwarfs and subdwarfs in the PG
(see \S~2.1), and (2) the crudeness with which \citet{fleming86} were
able to estimate $\Te$ values for each star using a heterogeneous set of
pre-CCD era spectrophotometry.  The method is one-dimensional since a
monotonic relation between $\mv$ and $\Te$ assumes basically that all
white dwarfs have the same mass.  Moreover, as we shall show, the PG
sample would need a substantial correction for incompleteness at the
redder colors (fainter $\mv$ bins).  The LF presented here is therefore
a crude, first estimate of a two-dimensional function. 

\subsection{The Mass Distribution} 

The entire mass distribution as a function of $\Te$ is shown in Figure
\ref{fg:f11}, together with evolutionary isochrones discussed
further below. Below $\Te\sim13,000$~K ($\log\Te=4.11$), the atmospheres
of DA stars become convective, and there is the suspicion that the high
masses inferred from spectroscopy below $\sim 12,000$~K are actually a
measure of the presence of helium brought to the surface by the hydrogen
convection zone \citep[][BSL]{bergeron90}.  As in BSL, our strategy is
thus to exclude all stars below 13,000 K from the following
discussion. The resulting gravity and mass distributions for the 297 DA
stars above 13,000~K are shown in Figure \ref{fg:f12}.  A bin size of
0.1 dex was used for the gravity distribution, which has a mean of
7.885.  As the mass distribution is the primary parameter of interest,
we do not discuss the gravity distribution further.

The mass distribution shown is constructed from a binning of the 297  
masses over the range 0.2 to 1.2~\msun, using a bin size of 0.025~\msun.
This is a significantly finer binning than the commonly employed
0.05~\msun\ used by many previous studies having smaller sample sizes.
To within the limits of poisson statistics, the complete PG mass
distribution can be well characterized by three primary components with
assumed Gaussian profiles. The primary peak is centered on 0.565~\msun,
with a FWHM of 0.188~\msun.  At the lowest masses there is a peak
centered on 0.403~\msun, with a FWHM of 0.055~\msun.  Finally, we
tentatively identify a broad, high-mass component centered on
0.780~\msun\ with a FWHM of 0.255~\msun.  This, unfortunately, appears
to overlap considerably in mass with the primary 0.6 \msun\ component,
{\it if} the assumption of a Gaussian shape is correct.  As we shall
discuss, this assumption does not predict the correct number of stars
with masses $>1.0$ \msun.  The respective contributions of the
0.4~\msun, the 0.6~\msun\ and the $>0.8$~\msun\ components are 8\%, 76\%
and 16\% of the unweighted sample.

The peak just below 0.6 \msun\ was first pointed out by \citet{ksw79},
and the sharpness of this peak is now well documented from several
modern studies already cited \citep{bsl,brb95, marsh97,vennes97b}.  In
the entire PG sample the peak is located at 0.572~\msun, significantly
less than our mean sample mass of 0.62~\msun.  (This is due to the fact
that there are more stars in the high mass component than in the low
mass component.)  The intrinsic width of this peak should be slightly
less than the 0.188~\msun\ FWHM, due to the broadening introduced by the
inherent uncertainties in the estimated masses.  However, the dominant
central component is effectively resolved.

Assuming that all of the stars in the central peak represent single
white dwarfs, and the number of undiscovered double degenerates is low,
then approximately 76\% of the PG white dwarfs fall into a fairly narrow
range of masses.  Considering the stars within one standard deviation of
the peak, the mean age and temperature of these stars is $7.5\times10^7$
years and 27,400~K, respectively.  The corresponding range of progenitor
main sequence masses of these stars is also fairly restricted.  Using
the revised initial-final mass relation of \citet{weidemann00} ,
this range of white dwarf masses would correspond to main sequence
progenitor stars with masses between approximately 1.0 to 2.5~\msun,
essentially spectral types G2 to A0.  If the initial-final mass relation
were strictly monotonic with no dispersion, the central peak should have
a sharp low-mass cutoff -- see \citet{wy89}.  The relative symmetry of
the wings suggests that there is appreciable dispersion.

The low mass peak consists of He-core white dwarfs, formed when the
He-core burning phase is truncated during a common-envelope phase in
close binary systems.  As predicted in BSL, these stars turn out to be
close binaries, usually double degenerate systems \citep{marsh95}.
These must be white dwarfs with cores composed of helium, whose
envelopes were removed in the red giant phase before the core mass
reached the 0.45--0.50~\msun\ necessary for core-helium ignition
\citet{ibenrenz84}.  Note that the minimum between the two peaks is in
this interval, as would be expected.  The lowest mass case is PG~1101+364
at 0.32 \msun.  In our analysis we implicitly assume that the
spectra of all He-core white dwarfs are dominated by a single luminous
degenerate star.  The existence of significant spectroscopic or
photometric contributions from a secondary degenerate star can bias our
mass and distance estimates for such stars.

There also exists a distinct high mass shoulder next to the central
peak, which we are able to characterize as a much broader Gaussian with
a centroid of 0.780~\msun\ and a FWHM of 0.255~\msun.  Computing means
for the stars with in $\pm 0.1$~\msun\ of the centroid yields
characteristic ages and temperatures of 1.1$\times$10$^8$ years and
25,500~K, considerably older and cooler then the central peak.  If these
result from single star evolution, the progenitors would have had a mean
mass of about 3.8~\msun, but there appear to be somewhat more of them
than modeling with a standard initial mass function would predict as
single-star evolution \citep{wy89}.  There may also exist a high mass
excess of stars between 1.0 and 1.2~\msun.  Several authors
\citep{vennes97b,vennes99,marsh97} have suggested the existence of one
or more high mass peaks in this range.  However, these were based
primarily on EUV selected samples of hot DA stars.  While there well may
be selection effects with these samples that are responsible for an
apparent excess of high mass stars, in our sample we can use the
Gaussian tail of the high mass shoulder to estimate the expected number
of stars in the range 1.0 to 1.2~\msun.  Extrapolating the high mass
shoulder indicates that there should be less than one star in our sample
with a mass this high, in contrast there are eight such stars.  The few
such objects expected from single-star evolution with masses above about
1.2 \msun\ might have cores composed of oxygen-neon-magnesium (O--Ne--Mg),
the primary products of the nucleosynthesis of $^{12}$C.

It has been been suggested by several authors, due to this apparent
excess of massive stars, that a substantial fraction of the
0.8--1.35~\msun\ result from close binary evolution, in particular
mergers white dwarfs of double helium (He--He), helium-carbon-oxygen
(He--CO), and double carbon-oxygen (CO--CO) core compositions.  This role
was in fact suggested by \citet{marsh97} and \citet{vennes99}, among
others.

The mass distribution discussed above, however, is biased substantially
in a magnitude-limited survey like PG, due to the corresponding
variation in the radius of the white dwarfs.  The search radius to a 
given limiting magnitude is proportional to $R$, so the search volume 
is proportional to $R^3$ \citep{shipman72,liebert95}. 
The search volumes available for each star differ for this and other 
reasons, but the $1/V_{\rm max}$ weighting corrects for all of these. 

Hence we show in Figure \ref{fg:f13} the mass distribution corrected for
the actual volumes searched.  This distribution should display the true
space densities of white dwarfs of various mass intervals.  For the
three mass components identified earlier, the total space densities of
DA stars with $\Te\gta\ 13,000$~K are 4.52$\pm$0.03$\times$10$^{-6}$
pc$^{-3}$ for the 0.4~\msun\ component, 1.183$\pm$0.75$\times$10$^{-4}$
pc$^{-3}$ for the 0.6~\msun\ ``peak'' component, and a surprising
3.53$\pm$0.22$\times$10$^{-5}$ pc$^{-3}$ for the $>0.8$~\msun\ massive
component.  The total space density of DA white dwarfs based on this PG
sample is 1.58$\times$10$^{-4}$ pc$^{-3}$, while the three mass
components account for 3\%, 75\%, and 22.0\% of the hot DA sample,
respectively.  Because of the absence of lower temperature, lower
luminosity stars, this total density is considerably lower than that
estimated in the previous subsection from the LF for $\mv <12.75$.

The shape of Figure \ref{fg:f13} suggests that any high mass
component may split into sharper components centered near 0.9~\msun\
and beyond 1.0~\msun.  The latter accounts for a space density of
1.02$\times$10$^{-5}$ pc$^{-3}$, or 6.4\% of the total DA white dwarf
density!  We have already noted that such a separate component of $>1$
\msun\ white dwarfs exists has again previously been identified in
their EUV-selected samples by such authors as \citet{marsh97} and
\citet{vennes99}.

These PG DA white dwarf density estimates are just lower limits,
however.  It is necessary to correct for the incompleteness of the
survey, for white dwarfs hidden by luminous nondegenerate companions,
for stars that are actually double degenerate systems, and for DB/DO
stars if one desires the total space density of hot white dwarfs.  Most
importantly, as we shall document in the next section, the consideration
of the space densities of different components of white dwarfs gives an
incomplete, and probably misleading, picture.  One must fold in the
evolution times for stars of different masses, in order to calculate the
formation rates of each group of stars.

\subsection{The Luminosity Function Revisited: The Three Mass 
Components} 

Since the 0.4~\msun, 0.6~\msun, and $>0.8$~\msun\ components may
represent, at least in part, separate paths of single and binary stellar
evolution through the white dwarf phase, it is appropriate to break up
the LF of Figure \ref{fg:f10} into (at least) three components.  This
largely removes the mass dependence on the LF, and leads to insight into
the selection effects due to magnitude and color (temperature) limits.
For now we will retain the $>0.8$~\msun\ component intact.  Small
numbers would make it difficult to subdivide the component. 

In Figure \ref{fg:f14}, as expected, it is seen that the
0.6 \msun\ peak contributes the most to the white dwarf space
density.  The $>0.8$~\msun\ component rises much more sharply at
lower luminosities.  While contributing very little at $\mv<10.25$,
near $\mv\sim13$ its contribution comes close to matching that of the
0.6~\msun\ LF.  Some understanding of this behavior comes from the
very different cooling rate for massive white dwarfs, compared to the
norm.  At highest luminosities, these cool more rapidly due to neutrino
losses from their dense cores.  At lower luminosities, however, they
cool much more slowly: they have much greater internal energy per unit
volume in the core, but smaller surface areas from which to radiate it
away.  So the massive stars evolve over this interval from cooling more
rapidly to cooling considerably more slowly than the 0.6~\msun\ peak
stars.  As to their identification in the PG Survey, their cooling times
offset the selection effect due to the smaller survey volume in which
they may be found.

Very contrasting in behavior is the distribution of the 0.4~\msun\
component.  This LF contributes down to $\mv\sim11$, but nothing
fainter than that magnitude.  These will linger as hot stars since their
low density interiors lose little energy to neutrinos.  A selection
effect, however, is very important: at $\mv>11$, they are
luminous due to their large radii, but the $\Te$ fall below the 10,000~K
limit due to the UV--excess selection of the PG Survey.  Their rapid 
evolution time offsets their larger survey volumes in the PG selection.

\subsection{The Mass and Luminosity Distributions vs. Temperature} 

Crossing the mass range in Figure \ref{fg:f11} are evolutionary
isochrones with cooling times of $\log\tau=5.5$--9.5 years obtained from
the carbon-core models of \citet[][solid lines]{wood95}.  This is not
appropriate for masses much below 0.5 \msun, and we will return to this
point shortly.  It is seen that completeness certainly stops near
log~$\Te$ of 4.0, where the 0.4 \msun\ stars are near $\mv\sim11$ but
the $>0.8$ \msun\ stars approach $\mv\sim13$, as we saw in Figure
\ref{fg:f14}.  The faster neutrino-cooling phase of the high mass stars
is brief, and these stars cool more slowly at cooling times 
$\tau>10^6$ years.  They cool much more slowly by $10^9$ years.  The 0.4
and 0.6 \msun\ show sharp peaks in two dimensions, while the
distribution of high mass stars is again more extended.

An alternative way of highlighting similar themes is to plot the
absolute visual magnitude $\mv$ vs.~$\log\Te$.  This is shown in Figure
\ref{fg:f15}. Diagonal solid curves from the upper left to the lower
right are lines of constant mass from the Wood models, from 0.4 \msun\
(top) to 1.2 \msun\ in intervals of 0.2 \msun, as labeled in the lower
left of the figure.  The dashed curves are evolutionary tracks from He
interiors of \citet{althaus01}, with the masses labeled in the upper
right of the figure.  Crossing the mass lines are isochrones with log
cooling times for 0.5 dex intervals from 5.5 to 9.0.  Note that the
helium track isochrones for 7.5 and earlier are too luminous to fall in
this plot.  Here the slope of the isochrones due to the slower cooling
of the massive stars turns nearly horizontal!  Incompleteness near $\Te
\sim10,000$~K clearly imposes a sharp cutoff in $\mv$ as a function of
mass.

\section{COMPLETENESS OF THE PG SAMPLE} 

PG was a photographic survey performed in the 1970s using the Palomar
46-cm Schmidt telescope.  The detector was baked IIa-O film.  The fields
were circular in shape, and overlapped by $<$10\%.  A double, offset
$m_u$ and $m_b$ exposure was made on each field.  Point sources were
scanned with a microdensitometer.  Selection of ``UV-excess'' candidates
was made on the basis of a photographic $m_b$ limiting magnitude (near
16.2) and a $m_u$--$m_b$ color cut. Photoelectric photometry was
obtained of a few standards in each PG field in order to calibrate to a
photometric system.  the accuracy of the $m_b$ values was believed to be
0.29 mag. The color cut was applied to identify candidates having
$m_u$--$m_b<-0.46$, but the dispersion in $m_u$--$m_b$ was modelled as a
Gaussian with $\sigma$ = 0.38 mag.  Star-count modelling tied to the
local standards showed the $m_b$ limit varied from field to field, and
such observations were necessary to determine it for each field.
Moreover, followup spectroscopy showed that the actual $m_u$--$m_b$ cut
also varied considerably from field to field, as was evident in
particular by large variation in the number of weak-lined main sequence
stars extracted as candidates, because of the large uncertainties, and
the realized $m_u$--$m_b$ cut varied appreciable from field to field.

The preceding paragraph is a brief summary of the details provided in
\citet{PG}, to give the reader a sense of how this three-decades-old
survey was carried out.  Some 46.8\% of the survey area was covered by
two exposures, and various tests suggested to the authors that the
overall completeness to the specified magnitude and color limits was
about 84\%, although others (see below) have argued that this number
is too optimistic.  In estimating the luminosity function,
\citet{fleming86} attempted a color incompleteness correction which
increased with $\mv$ (see their Table 2). We shall argue below,
however, that this correction was probably too small for the fainter
$\mv$ bins.

In the following decades, deeper surveys of fields overlapping the PG
were performed, so that subsequent authors could evaluate the PG
completeness.  The Edinburgh-Cape Blue Object Survey also targetted the
North Galactic Cap at Galactic latitude $|b|>30^o$, to a fainter
limiting magnitude of $m_b=16.5$, and to somewhat redder stars.  
\citet{edincape97} concluded that the completeness of
the PG averaged 84\% relative to other surveys and compilations, which
could themselves be incomplete.  The area of overlap is not stated in
the paper.  These authors appeared to dispute the much less optimistic
conclusions of \citet{gold92}, who claimed a surface density of quasars
a factor of 3.4 times higher than that found in the PG.  The Goldschmidt
group appeared to use an appreciably-fainter $m_b$ limit, which could
make a large difference since the surface density of UV-excess quasars
is known to rise sharply at $B\sim16-17$.

\citet{darling94} -- see also \citet{wegner94} -- did a similar comparison
of the Kiso Survey \citep{nogu80,kondo84} with the PG.  The Kiso Survey
used $m_u$ and $m_g$ photographic colors, with 600 square degrees of
overlap.  \citet{darling94} estimated the completeness of PG to be 57.5\%
for white dwarfs to the stated magnitude limits of the PG (though only
50\% for quasars).  Using his spectroscopically-identified sample of hot
white dwarfs, \citet{darling94} constructed a LF from the Kiso or KUV
sample, shown in Figure \ref{fg:f10}.  \citet{wegner94} stated that
``On the whole, the luminosity function derived compares relatively well
with that from the PG Survey \citep{fleming86}.''  In fact, the KUV LF
has significantly more stars than PG in the bins $\mv$ 10.5 to 13.0,
with the PG appearing to become incomplete by a factor of 4 in the 12.0
and the two fainter magnitude bins.

It has to be noted that, while spectra had been obtained to classify the
DA white dwarfs of the \citet{darling94} KUV sample, the assignment of
$\mv$ values for individual stars was not based on the spectroscopic
line-fitting method utilized here.  Rather, an $\mv$ vs. $m_u-m_g$
relation was determined from $B$--$V$ photometry of a subset of the
sample -- the $\mv$ - color index or $\mv$(CI) relation.  The $B$--$V$
vs. $m_u-m_g$ relation had a very large dispersion -- see Figure 4.1 of
\citet{darling94} -- and necessarily so did the $\mv$ estimates.

We have been able to assess the dispersion and look for systematic
error in the \citet{darling94} $\mv$ estimates using KUV stars for
which accurate CCD spectra are available -- largely due to observation
by one of us (P.B.), using the same Steward 2.3m telescope and
spectrograph.  Excluding four DA+M stars for which the CI might be
corrupted by the companion, 67 KUV stars can be used to compare the
spectroscopic line-fitting luminosities $\mv$(sp) with
$\mv$(CI). The mean difference in $\mv$(sp)--$\mv$(CI) is
$-0.526$, and a linear fit yields a relation of the form $\mv({\rm
sp})=-3.118 + 0.242\,\mv$(CI). Part of the growing difference between
the KUV and PG LFs as $f(\mv)$ in Figure \ref{fg:f10} is
understandable from the systematic offset in $\mv$ between the two
sets of determinations.  However, this offset varies from from about $-1$
at $\mv$(sp)= +9 to 0 at $\mv$(sp)= +13.  Since the KUV survey includes
a redder color limit than PG, it is clear that the latter becomes 
increasingly incomplete for the fainter magnitudes (redder colors), 
due to the large dispersion in the $m_u$--$m_b$ color measurements. 
In the following section, we present an additional argument.  However, 
the offset in $\mv$ between the two surveys makes the KUV sample 
difficult to deal with for deriving a completeness correction for 
the bluer, hotter PG stars. 

The EUVE and ROSAT all-sky surveys detected only hotter white dwarfs, 
generally with $\Te >25,000$~K.  We have found that 18 EUVE sources, 
listed in \citet{vennes97b} and \citet{vennes99} (but including 
ROSAT detections), were also found in PG.  The EUVE and ROSAT 
surveys detected many white dwarfs with $B$ magnitudes fainter than 
the PG limit, which makes this a good source for testing the 
incompleteness in PG for hot white dwarfs -- i.e. too blue for the 
$m_u$--$m_b$ color dispersion to be a significant problem.  Rather, 
this is a cleaner test of the magnitude limit of the survey.  Note 
that the application of this test does not require the assumption 
that the EUV surveys are 100\% complete in detecting white dwarfs 
to a particular flux level.  

The 18 EUV-detected sources listed in the above references have PG
(1950) designations of 0102+095, 0136+251, 0216+143, 0937+505, 1026+453,
1033+464, 1040+451, 1041+580, 1057+719, 1109+244, 1125-025, 1145+187,
1234+481, 1254+223, 1403-077, 1415+132, 1636+351, and 2357+296.
However, 12 EUV-detected stars, believed to be within a PG field as
listed in the survey paper, and measured to have $B$ magnitudes within
the $m_b$ color limit listed for that field, do not appear in PG.
(Other EUV detected stars having V magnitudes from \citet{vennes97b}
fainter than the magnitude limit of the overlapping PG field are
ignored.)  The 12 include 5 -- PG 0232+035 (Feige 24), 1123+189,
1232+238, 1415+132, and 2309+105 -- which {\it are} in the PG, but in
fields which for reasons of quality were not retained in the final
sample of the catalog.  We omit 0845+488J, as well, since it has a
luminous A companion (HD~74389A) that would prevent its discovery in a
UV-excess survey.  The remaining six undetected stars, named by the
J2000 coordinates used in the EUV survey publications, are 1016-053J,
1032+534J, 1043+490J, 1426+500J, 1529+486J, and 1650+406J.

We thus find that 18 of 24 EUV-selected stars in PG fields are in the
complete PG sample, or 75\%.  This falls between the estimates of
\citet[][58\% for white dwarfs]{darling94} and \citet[][84\% for
completeness to the stated magnitude limit]{edincape97}.  The latter
value agrees with the assessment in the PG paper itself.  Our judgment
is to adopt this intermediate value of 75\%, given that the
\citet{darling94} result was probably influenced by the color
incompleteness, a problem which the PG project clearly underestimated.
At the same time, the evidence from the EUV comparison is that the PG
estimate was too optimistic.  However, we have to regard this
completeness correction as uncertain by at least 15\%.

There is one remaining worry about the application of this test: Four of
the six ``missed'' EUV-selected stars have V magnitudes of 13-14.  These
magnitudes are near the brightness limit of the film.  The possibility
exists that the photographic $m_u$ or $m_b$ could be reaching saturation
at these brightnesses, so that the color difference was corrupted.  This
is difficult to quantify, since there is no hard magnitude limit.  It 
is possible that apparently-bright stars that have been missed could
have added disproportionally more to the $1/V_{\rm max}$ density.

\section{THE FORMATION RATE OF WHITE DWARFS}

\subsection{Formation Rate of the DA White Dwarfs in PG} 

The more complete picture of the DA white dwarfs is to take the space
densities and consider their evolution times as functions of both mass
and luminosity.  In Figure \ref{fg:f16} the number density of
white dwarfs weighted by $1/V_{\rm max}$ is plotted for 0.5 dex
intervals of the log of the cooling time, for the three mass components
shown in Figure \ref{fg:f14}.  The evolution rates per unit time
interval are listed in Table~3.  The local formation rate of white
dwarfs is not expected to have varied rapidly over the last Gyr or so.
We thus expected the evolution rates for each of the three mass
components to be roughly constant, as long as there are enough stars per
bin and the samples are complete. 

For the 0.6~\msun\ peak component, the rates are similar for the first
four bins through log~$\tau<$ about 7.75.  This corresponds roughly to a
$\Te$ near 23,000~K and $\mv\sim$10.5. For the oldest four bins,
however, the rates decline rapidly.  This indicates the onset of
incompleteness, due in all probability to the increasingly red
$m_u$--$m_b$ color of the stars.  The redder a star is, the more likely
it was not found in the PG.  The dispersion of photographic $m_u$--$m_b$
must be large indeed for any DA stars approaching 25,000~K to be
missed. This decline in the rates tracks the progressive decline of the
overall LF compared with that of \citet{darling94} for the fainter $\mv$
bins.  A rough weighted mean of the first three bins of Table~3
indicates that the formation rate of dominant 0.6~\msun\ component DA
white dwarfs is thus most likely around
4.5$\pm$0.4$\times$10$^{-13}$~pc$^{-3}$ yr$^{-1}$.  The error estimate is
based on the dispersion of the first three bins.  A single star with a
CO core usually dominates the light, but a fraction of these may be
double degenerates, for which no enhancement to the rate is
applied. Moreover, no correction has yet been applied for the likely
incompleteness due to the PG magnitude limit discussed by
\citet{darling94} and \citet{edincape97}.

The other two components contribute appreciably to the overall formation
rate of white dwarfs.  Inspection of Figure \ref{fg:f15} reminds us,
however, that the onset of incompleteness due to $m_u$--$m_b$ color occurs
at different $\mv$ values.  We can make a rough estimate of the contributions
of the other two components, but the smaller sample sizes make these even
less rigorous.

The low mass (0.4~\msun) component reaches a similar color (25,000~K) at
$\mv$ near +9 and log~$\tau$ of 7.  Based on $\sim$10 stars, the rough
formation rate is near 0.4$\times$10$^{-13}$ in the same units.  

Since we intend to count the formation rate of all white dwarfs, in
binaries as well as single stars, our calculated rate requires
enhancement because most of the low mass stars are really double.
\citet{marsh95} found most of the low mass PG stars to be radial
velocity variables, though \citet{maxted98} investigated two that
appeared not to vary.  The SPY project is a large, general observational
program to find close binaries from several white dwarf surveys, mainly
as potential progenitors for SNIa \citep[]{napi01,koester01}.  Detailed
statistics are not yet available, but the yield of radial velocity
binaries from stars with low mass estimates appears to be high.  When
the primary star (in apparent magnitude) is a helium white dwarf, the
companion can have a core composed of He or CO.  Admittedly,
\citet{zb92} found that a few of the low mass PG white dwarfs harbor low
mass nondegenerate companions.  These should also be close, probably
pre-cataclysmic binaries with only one helium white dwarf.

Our conclusion is that, to account for the degenerate companions to the
observed low mass white dwarfs, the overall formation rate should be
0.6-0.7$\times$10$^{-13}$ in the same units. Note that this may include
some more massive companions with CO cores.  Thus the properly-weighted
contribution of the low mass stars is more than an order of magnitude
higher than indicated by the space density alone, and accounts for 
$\sim$10\% of all DA white dwarfs. 

There is another reason why the discovery of low mass probable binary
white dwarfs may be incomplete: they may be so low in mass, their
spectral lines are too narrow for them to be readily recognized as white
dwarfs.  The discoveries by (1) \citet{heber03} that HD~188112, a
``bright $V=10$ nearby B-type star'' (to quote these authors), is in
reality a very low mass (0.24~\msun) white dwarf with a helium core, and
(2) of several more in the early white dwarf samples from the SDSS
\citep[]{kleinman04,liebert04} with masses as low as 0.19~\msun,
suggests that the low mass white dwarf component is incomplete in its
mass range.  The lowest mass white dwarf in the present sample is
0.32~\msun.  No additional correction to the rate for this kind of
incompleteness can yet be made, but the ultra-low mass stars can be
found efficiently in the SDSS samples.

For the high mass component, the stars remain sufficiently blue
(25,000~K) to about $\mv$ near +11 and log~$\tau\sim$ 8.0.  Again, with
$\sim$10 stars retained that are sufficiently blue, a rough formation
rate of 0.9$\times$10$^{-13}$ is inferred.  This is 20\% of the rate of
the peak 0.6~\msun\ component, a factor of two smaller than the relative
contribution from the space density alone.  However, the contribution of
the subset with $\ge$1~\msun\ drops to several per cent from the 14\%
contribution to the space density.  There are not enough stars to
determine a separate, accurate formation rate for this very massive
subset.

In the next section, the likely contribution of binary evolution to the
component of massive white dwarfs will be evaluated.  It is
inappropriate to increment the formation rate for binaries within this
component, as binary evolution would generally result in a merger
leaving a single star of high mass.  Very few double degenerate systems
with very high combined masses have been found so far in the SPY project
\citep[]{karl03}.
 
No correction to the contribution of the massive star rate for those in
binary systems is appropriate here.  Binary evolution may be responsible
for the some of the massive stars -- presumably by mergers of white
dwarfs of more ordinary mass -- as argued previously by from their EUV
and soft X-ray-selected hot white dwarfs.  If the high mass component is
thus a mix of these two origins, very few if any will remain binary, and
no enhancement to their rate is appropriate.

Proper weighted by the apparent magnitude limit (radius), and by 
cooling time, the low and high mass components account for 15\% and 
20\% of all hot DA white dwarfs.  

Thus, our estimate of the total formation rate of hot DA white dwarfs is
about 6$\times$10$^{-13}$~pc$^{-3}$ yr$^{-1}$.  Of these 10\% are of 
low mass (or have at least one binary component of low mass), and a 
striking 15\% are more massive than 0.8~\msun.  The total estimate is 
likely uncertain by at least 10\% based on the scatter in the rate 
determinations.  

\subsection{Inclusion of DB and DO White Dwarfs} 

To get the total formation rate for all hot PG white dwarfs, one must
also include the 51 DB and 9 DO stars found in complete sample.  A
similar analysis of these stars has not yet been completed.  We do know
that the temperature (and age) distributions of the DB-DO and DA
sequences differ in that the DB-DO do not include any stars between
30,000~K and 45,000~K (the so-called ``DB gap'').  The DA sequence
overlaps the DB-DO sequence at other temperatures.  For now, we can only
assume that, on average, the same $V_{\rm max}$ weighting distribution
applies for these stars.  This would give us an increment of 60/347 or
17\%.  If the spectroscopic analysis of a large sample of DB stars by
\citet{beauchamp96a} and \citet{beauchamp96b} is correct, nearly all of
the stars with helium atmospheres are part of the 0.6 \msun\ component.
It should be noted, however, that the more modest study of DB stars by
\citet{koester01} found systematically higher masses.  Anyhow, the
overall formation rate is lifted to 7$\times$10$^{-13}$~pc$^{-3}$
yr$^{-1}$.  It is impossible to assess how much is the uncertainty in
this non-DA rate increment, though it is a relatively minor effect.

\subsection{Catalog Incompleteness Due to the Magnitude Limit} 

Our comparison with lists of EUV detections discussed earlier favors
75\% as the best educated guess as to completeness to the stated
magnitude limits of individual PG fields.  This applies for only white
dwarfs hotter than about 25,000~K for reasons discussed earlier.  With
this assumption, the overall PG rate is multiplied by 1/0.75, to give a
corrected rate of 0.93$\times$10$^{-12}$~pc$^{-3}$ yr$^{-1}$.  The
optimistic assessment of 82\% catalog completeness from
\citet{edincape97} and the original PG paper would yield
0.85$\times$10$^{-12}$~pc$^{-3}$ yr$^{-1}$, while the pessimistic 
\citet{darling94} estimate (58\% for white dwarfs) would yield 1.2 
in the same units.  The uncertainty in the completeness estimate is 
at least 15\%.  

\subsection{Incompleteness due to White Dwarfs Lost from PG due to 
Luminous Companions} 

A further contribution to the formation rate of white dwarfs comes from
undetected white dwarfs in binary systems containing luminous
nondegenerate companions.  Such systems, typified by Sirius and Sirius
B, are often called Sirius-like systems.  However, unresolved main
sequence companions as cool as K stars may prevent selection of the
white dwarf by UV excess. The relative frequency of these systems is
highly uncertain since the white dwarf components are effectively
obscured at optical wavelengths by the luminosity of the companion.  In
the past they have mainly come to light through radial velocity
variations of the primary or in some cases, were accidentally
discovered as wide binary common proper motion pairs.  Recently,
however, observations in the vacuum UV and the EUV have led to
the discovery of many more such systems.  At these short wavelengths,
even moderately hot white dwarfs ($\mv<12$) can dominate the
composite energy distribution.  

A survey of existing literature (by JBH) indicates that there are
approximately 41 Sirius-like systems with white dwarf companion of types
B through K, most having been discovered at EUV wavelengths.  A
relatively firm lower limit to the relative number of such systems can
be obtained from the known white dwarfs within 20 pc of the Sun
\citep[]{holberg02}.  Of the 109 white dwarfs in this sample, 25\% are
in binary systems, mainly with M dwarfs.  All four Sirius-like systems
in the sample lie within 14 pc of the sun.  This can be taken as a
reasonable lower limit of $\sim4$\% on the relative frequency of such
systems.

The actual number is expected to be significantly higher, however.  Some
evidence that this is indeed the case comes from the $\it FAUST$ UV
camera flown on the space shuttle Atlas-1 mission \citep[]{bowyer93}.
Detailed studies of 14 selected $\it FAUST$ fields
\citep[]{fromiggini03} detected some 14 main sequence stars with strong
evidence of UV excesses, where only 3 to 5 would have been expected
based on the number of EUV detections.  The EUVE and ROSAT EUV surveys
doubtlessly failed to detect many Sirius-like white dwarfs, due to (1)
the distance limitations imposed by the absorption in the interstellar
medium, (2) effective temperature biases due to a lower limit of
$\Te=25,000$~K for detecting any white dwarf, and (3) the rapid increase
of EUV opacity in most DA stars above $\Te=50,000$~K.  Finally,
consideration of the cumulative log N - log distance distribution for 39
Sirius-like systems, having firm distance estimates, gives a slope of
only +1, where a slope of +3 slope is expected for a reasonably complete
sample.

In summary, an uncertain increment of around 10\% to account for white
dwarfs hidden by luminous companions appears appropriate.  We shall
include this tacitly in the final estimate.  A program has been
initiated using the $\it GALEX$ (Galaxy Evolution Explorer) mission to
produce hopefully a much better estimate of the fraction of hot white
dwarfs hidden by companions.

\subsection{The Total White Dwarf Formation Rate} 

In conclusion the total, corrected rate of recent white dwarf formation
appears to be close to 1$\times$10$^{-12}$~pc$^{-3}$ yr$^{-1}$.
Our best educated guess is that this is uncertain by $\pm$0.25 in 
the above units, but this is obviously poorly constrained.  

It may be worthwhile to compare this rate applicable to white dwarf
formation in the last 10$^8$ years or so to estimates of the average
formation rate over the lifetime of the Galactic disk.  A rough estimate
may be easily obtained.  The local density of all white dwarfs is
arguably complete to a distance of about 13~pc, based on the cumulative
distribution function \citep[]{holberg02}, and is
5.0$\pm$0.7$\times$10$^{-3}$ pc$^3$.  Since these and other cool white
dwarfs fit cooling ages no older than 8 Gyr (BLR), the average white
dwarf formation rate averaged over the lifetime of the disk is about
6$\times$10$^{-13}$~pc$^{-3}$ yr$^{-1}$.  It is not surprising that the
current formation rate is nearly twice as high, since new white dwarfs
are drawn from progenitors of decreasing average mass.

\subsection{Other Estimates of the WD Formation Rate} 

\citet{vennes97b} estimated a DA white dwarf formation rate of
0.7-1.0$\times$10$^{-12}$~pc$^{-3}$ yr$^{-1}$ based on a sample of over
100 hot white dwarfs detected in the EUVE.  Objects fall between
25-75,000~K and most have $\tau < 30$ Myr.  It may be argued that EUV
selection may be hampered by neutral hydrogen absorption in the local
ISM or heavy element absorption in the white dwarf photosphere
\citep{finley97,marsh97}.  Both effects might exclude some white dwarfs
from detection, and bias an EUV sample toward higher masses where
gravitational diffusion could more likely be effective.  However,
\citet{vennes97b} found that the estimated density based on samples of
objects closer than 80-100 pc is only marginally dependent on Galactic
coordinates.  A cumulative distribution vs. distance of 110 DA stars is
fairly well reproduced by a $d^3$ slope up to 80-90 pc.

Eight stars from this PG sample have $\Te>25,000$~K and estimated
distances $<90$ pc.  Of these, six are detected by the EUVE.  Moreover,
the \citet{vennes97b} formation rate is in fairly good agreement with
that reported here after correction of the PG DA sample for catalog
incompleteness and luminous binary companions.

The older \citet{weidemann91} estimate of 2.3$\times$10$^{-12}$ has
often been cited by those comparing the formation rate of planetary
nebulae (PN).  However, this result presented in a conference proceeding
is not based on a study of a hot white dwarf sample.  Rather, it is
based on the star, white dwarf formation model over the lifetime of the
Galactic disk by \citet{yuan92}, mentioned earlier.  Moreover, the rate
would be only 60\% of this value if a 67\% correction for incompleteness
due to luminous companions were not included.  The rate does agree with
recent estimate of the PN formation rate, as we will discuss.  If our
binary correction of 10\% were applied, the Weidemann estimate is
actually in fair agreement with ours and \citet{vennes97b} estimates.

\citet{darling94} did not calculate an overall formation rate of DA 
white dwarfs.  It may be possible to correct the $\mv$ bias and make 
such an estimate.  It would likely exceed our uncorrected PG rate, 
but not necessarily the corrected rate.  

\subsection{Comparison with PN Formation Rate}

To determine an accurate formation rate of PN in the Galaxy has been a
notoriously difficult problem. \citet{pottasch96} notes that published
values differ from 0.4-8$\times$10$^{-12}$ pc$^{-3}$ yr$^{-1}$. There are
two main reasons why this rate is difficult to determine:

(1) Distance estimates for Galactic PN are obtained by a variety of 
methods, but traditional distance scales vary over a factor of 2, 
which translates to a factor of 8 in space density. 

(2) Lifetimes of the PN phase are very uncertain.  For example, in
individual cases, there is often disagreement by an order of magnitude
between a nebular kinematic age (several $\times$10$^4$) and the 
evolutionary age of the central star (several $\times$10$^5$) based on 
its mass and position in the HR Diagram.  

The best estimates of the space density of PN utilize a local sample,
where the distances may be estimated by more robust techniques, such as
a trigonometric parallax.  However, this still requires an accurate
census of sources close to the Sun, as well as distances and post-AGB
ages.  Two such recent studies may be the most reliable attempts for 
Galactic PN. 

For a restricted cylinder of radius 1 kpc and vertical length
0.64 kpc, \citet{phillips02} gets
2.1$\times$10$^{-12}$ pc$^{-3}$ yr$^{-1}$.  A difficulty with the
comparison to white dwarfs, even with a ``local'' sample, is that the
use of a much larger sample size for PNs like the above is inevitable,
and leads to a considerable dependence on the assumed vertical ($z$) scale
height. \citet{pottasch96} gave an estimate of the local PN formation rate
based on distances to 50 nearby PN determined individually by a variety
of methods.  Using a 0.7 kpc cylinder around the sun, his value is
3$\times$10$^{-12}$ pc$^{-3}$ yr$^{-1}$. 

Many researchers might conclude that estimates of the formation rate of
WD and PN differing by only a factor of 1.8 to 2.6 (for each of the
above estimates) are not in serious disagreement, given the many
uncertainties.  On the other hand, the PN are only a subset of the
objects becoming WD.  The PN phase probably must have been preceded by a
full AGB phase.  Stars which for any reason have lost much of their
hydrogen envelopes in their earlier red giant evolution may evolve off
the AGB at a lower luminosity and core mass.  These may have post-AGB
evolutionary times too slow to ionize the ejected PN gas before it is
dispersed.  There are more ``UV-bright'' post-AGB stars observed in
globular clusters than there are PN.  Hence, WDs emerging from the
oldest and lowest mass progenitors as single stars may not produce PN.
Some horizontal branch stars have lost so much hydrogen envelope that
they do not enter the AGB phase at all, and almost certainly do not
produce PN.  \citet{heber86} has estimated that about 2\% of WD
progenitors go through only such a subdwarf B phase of core-helium
burning.  While the majority of WD progenitors in the disk population
probably do pass through a full AGB phase, it is not clear how close to
100\% that fraction is.  Especially intriguing is the growing evidence
that a large fraction of PN central stars are binary
\citep[]{demarco04}, suggesting to some that single stars might not
produce PN at all!

In summary, it is not clear how serious is the lingering disagreement
between our estimate of recent WD formation rate and the recent,
probably best estimates of the PN rate.

\section{ORIGIN OF THE MASSIVE WHITE DWARF COMPONENT}

The 15\% contribution of stars more massive than 0.8~\msun\ to the rate
of formation of all DA white dwarfs would seem surprisingly high, based
on most previous studies.  We have already pointed out the suggestions
by \citet{marsh97} and \citet{vennes99} -- and even earlier by
\citet{schmidt92} and BSL -- that binary mergers may account for a
substantial fraction of these stars.  To our knowledge, the hypothesis
of a binary contribution has never been evaluated quantitatively.

To assess this possibility, one must first know the expected
contribution to the overall DA white dwarf formation rate from single,
presumably-massive progenitors.  In general, single white dwarfs of
$<$0.8~\msun\ are expected to evolve from $\ge4$ \msun\ progenitors,
which are rare compared to less massive stars.  The only theoretical
calculation of the expected mass distribution from single stars we are
aware of is the dissertation work of \citet{yuan92} -- see also
\citet{wy89}.  Here the overall white dwarf formation rate, mass and
luminosity distributions over the lifetime of the Galactic disk were
calculated, based on a stellar initial mass function (IMF), a star
formation rate, and an initial/final mass relation.  A graphical display
of one version of the predicted mass function is shown in Figure 11 of
\citet{yuan92}.  This particular choice assumes (1) a constant star
formation rate, (2) a Scalo IMF, (3) white dwarf evolution times from
\citet{wood90}, (4) a 12 Gyr disk age, and (5) the initial/final mass
relation of \citet{weidemann87}.  These choices still appear reasonable
today, but also the shape of the derived mass distribution appears from
the plots in the paper to be pretty insensitive to reasonable variations
of these five choices.

We have performed a graphical measurement of areas under the
``unfolded'' mass distribution curve in this figure.  ``Unfolded''
means no allowance was made for dispersion in the assumed monotonic
initial/final mass relation, an irrelevant consideration here. The
curve spanning 0.5--1.2~\msun\ was divided into ten arbitrarily chosen
mass intervals and linear segments were drawn across each.  The mass
intervals 0.5--0.55, 0.55--0.575, 0.575--0.60, 0.60--0.65, 0.65--0.7,
0.7--0.75, 0.75--0.8, 0.8--0.9, 0.9--1.0, and 1.0--1.2~\msun. Finer
intervals were chosen where the curve was changing slope the most.
Areas were measured under each chord to assess its contribution to the
overall predicted mass.  For the components $<$ and $>0.8$ \msun,
respectively, the areas (arbitrary units) are 172 and 6 units.  For the
$> 0.8$ \msun\ component, the subset $>1.0$ \msun\ is 2 units.

The above exercise indicates that the \citet{yuan92} calculation
predicts that the $>0.8$ \msun\ component forms only 3.5\% as many stars
as the $<$0.8~\msun\ ``peak'' component. The low mass component produced
from binary evolution is not part of this calculation. This compares
with 20\% as many high mass stars as ``peak'' stars from the rates of
the previous section.  The \citet{yuan92} results are given in units of
pc$^{-3}$, and hence allow for the magnitude limit correction.  The use
of cooling models from \citet{wood90} indicates that evolution times
have also been allowed for.  The only difference here is that the
calculated mass distribution includes white dwarfs with luminosities
down to below 10$^{-4}$~\lsun. 

The result of the calculation is that this model for white dwarf
formation from single star evolution predicts only 17.5\% of the
observed rate of formation of the massive white dwarfs.  Binary
evolution is required for 82.5\%!  We argue in Appendix~A, however, that
a larger fraction of the very hot, EUV-selected white dwarfs might come
from single-star evolution due to the associations of massive stars
distributed around the Sun nearer the Galactic plane.

\section{COMMENTS ON INDIVIDUAL OBJECTS} 

\subsection{Magnetic White Dwarfs} 

This sample includes 10 white dwarfs with detected magnetic fields. 
Eight of these have an estimate of the mass, from methods discussed 
in \citet{liebert03}.  These authors use this and other samples to 
document an increasingly-strong case that the average mass of magnetic 
white dwarfs is substantially higher than peak mass.  The mean mass of 
the eight PG stars is 0.93~\msun\ -- including PG~1658$+$441 at
1.31 \msun\ \citep{schmidt92}, the highest determination in the 
sample.  The radius bias discussed in \S~3 leads to the conclusion 
in the cited paper that the true frequency of magnetic white dwarfs 
with fields exceeding about 2 megagauss may 
exceed 10\% of all white dwarfs.  This is before allowing for more
difficult-to-detect fields less than this value. 

\subsection{Peculiar DAB and Composite Objects} 

Several stars with basically hydrogen-rich atmospheres but showing 
helium lines are found in the survey, usually with temperatures in 
the 30,000--45,000~K range of the so-called ``DB gap''
\citep{liebert86}.  GD~323 (PG~1302$+$597) is a peculiar DAB star at about 
30,000~K whose spectrum could be fit with neither a homogeneous, mixed 
H/He model.  A more promising, but unproven possibility is that its H 
and He layers are spatially stratified in some manner \citep{liebert84}.  
Since the parameter determinations for this star are not accurate, we 
have little choice but to omit it from the complete PG sample.  

PG~1305$-$017 and PG~1210$+$533 are peculiar DAO stars
\citet{bergeron94}.  The former is best fit with a stratified model at
$\Te$ near 44,000~K.  The latter ($\sim$45,000~K) shows variability in
the strengths of the He~I and He~II lines, perhaps due to spatial
variation in the H/He ratio.  Most DAO stars are best fit with
homogeneous atmospheres of mixed H,He composition, and are not known to
be spectrum variables \citep{bergeron94}.

The hydrogen-rich PG~1603$+$432 near 37,000~K was recently shown by
\citet{vennes04} to have a He~II 1085~\AA\ line in spectrum obtained with
the {\it Far Ultraviolet Spectroscopic Explorer}.  These authors
predicted that He~I 4471~\AA\ should be detected in a spectrum of high
signal-to-noise ratio.  These authors pointed out that the line may have
been marginally detected in a spectrum published by \citet{bergeron94}.
Indeed, we have obtained a second, better spectrum, and the He~I line
indeed seems to be there.  Both the old and new spectra of this star are
shown in Figure \ref{fg:f17}.  However, the best fit to the far-UV
spectrum that \citet{vennes04} obtained was with a homogeneous
atmosphere with He/H = 0.01.  This predicts a He~I 4471~\AA\ line that is
distinctly stronger than observed in both of these spectra.  This star
may thus be a spectrum variable, or require a more complicated H,He
layering.

The DAB star PG~1115$+$166, as mentioned in \S~2.3, requires a quite
different explanation.  \citet{bergeron02} showed that the spectrum is
best explained as a double degenerate where one component is a DA star
with $\Te$ of 22,000~K, accompanied by a DB star near $\Te$ of 16,210~K.
In a similar time frame, \citet{maxted02} found this to be a 30.09-day
period double degenerate system. 

\subsection{ZZ Ceti Stars} 

The PG sample extends to low enough effective temperatures to cover the
range where ZZ Ceti pulsators are found, between about $\Te=12,500$~K
and 11,100~K according to the detailed study of \citet{bergeron95}.  At
the beginning of our project, the PG sample included 9 previously known
ZZ Ceti stars, all of which were found within the empirical instability
strip. As part of our survey, PG 1349$+$552 (LP 133$-$144) was also
found to lie within the strip, and further high-speed photometric
observations by \citet{bergeron04} confirmed that it was indeed a new
variable white dwarf. Our results for the PG survey are thus consistent
with the conclusions of \citet[][and references therein]{bergeron04}
that the empirical instability strip contains no non-variable stars, in
sharp contrast with the results of \citet{mukadam04} who claim, based on
the analysis of the DA stars in the Sloan Digital Sky Survey, that the
ZZ Ceti instability strip contains a significant fraction of
non-variable stars.

\subsection{Two Unusual, Very Hot, Hydrogen-rich White Dwarfs} 

We could not end a section on the ``unusual'' objects in PG without 
mention of what may be the two hottest DA stars in the sample: 

BE~UMa (PG1155+492) is the only pre-cataclymic binary we are aware of in
the sample.  It consists of a DAO or high-gravity sdO paired with a K
star at a 2.2-hour period \citep[]{ferguson99} -- and references
therein.  The reprocessing spectrum due to radiation intercepted by the
facing side of the secondary is one of the most impressive known.  Of
the known sample of such binaries, this is one of the most recent to
have emerged from the common envelope evolution phase, which may account
for why the K star is oversized compared with main sequence stars of
similar temperature.  The $\Te$ and $\logg$ values are not considered
accurate enough in include it in the PG analysis here.

The planetary nebulae EGB~6 = PG~0950+139 is a unique, unsolved problem
of stellar astrophysics.  The central star has a poor $\Te$ estimate of
108,000~K, though it is still included in Table~1 and our analysis.
There is an old, low-surface-brightness PN shell \citep[]{egb84}, not
unexpected for a central star with these parameters.  What is unique
about the object, however, is that there is a $\it very$ compact
component to the PN \citep[]{liebert89}, very uncharacteristic of a
central star with these parameters.  Several arguments constrain the
dimensions of this extremely-dense, nebular component to the order of 10
a.u.  An unusual hypothesis to explain the object -- the ablation of a
Jovian planet by its very hot star \citep[]{dl89} -- is apparently ruled
out by imaging and spectra taken with the {\it Hubble Space Telescope}
\citep{bond93}.  The mystery remains.

\section{SUMMARY AND FUTURE ENDEAVORS}

While a greater appreciation of the incompleteness of the PG Survey --
particularly at lower temperatures -- over that envisioned earlier has
hampered our efforts, we believe that this thorough study of the sample
using homogeneous observations and improved theory has led to some
interesting, even unexpected conclusions.  When the sample is normalized
to the same volume, there are nearly two thirds as many high mass DA
stars (39.1\% of the total) with $\ge$0.8~\msun, as are in the peak
(0.46--0.8~\msun) component (60.1\%).  Yet, the low mass component -- so
prominent in the plot of the unweighted mass distribution -- accounts
for only 0.8\% of the space density of hot DA stars.

While the accounting for white dwarf radii has a big impact on a
apparent-magnitude limited sample, so does consideration of the
evolutionary times also as a function of mass.  Thus, the peak, low and
high mass components contribute 75\%, 10\%, and 15\% to the overall
formation rate of DA white dwarfs, respectively.  The calculation
discussed in \S~6 appears to require that $\ge$80\% of the massive white
dwarfs come from binary star evolution, presumably mergers of two white
dwarfs of smaller mass.  To our knowledge, neither the observed
formation rates as a function of mass components, nor the relative
contribution of binaries to the high mass component, have been
rigorously evaluated previously.  

When the overall formation rate is corrected for incompleteness of the
catalog (as best as we are able to assess it), for DB-DO stars, for
those hidden by luminous non-degenerate companions, and for those white
dwarfs which are likely to be double degenerates, we estimate a total,
recent formation rate of white dwarfs in the local Galactic disk of
1$\pm$0.25$\times$10$^{-12}$~pc$^{-3}$ yr$^{-1}$.  It is difficult to
know if this error estimate encompasses all possible systematic errors.
Agreement with previous determinations for white dwarfs is good.  The
best estimates of the formation rate of planetary nebulae still appear
to be around twice as high.  Although less of a discrepancy than
appeared to be the case at the time of the publication of
\citet{fleming86}, this may still be somewhat worrisome, since not all
white dwarfs pass through a planetary nebula phase.  A rigorous, similar
analysis for DB stars is also needed.

In the near future, larger samples with better completeness assessment 
and homogeneous observations can be brought to bear on this problem.  
The SPY project, though its primary purpose is to search for double 
degenerate systems using radial velocities, should produce robust 
determinations of the white dwarf parameters for ~1500 stars
\citet{koester01}.  Data from the Sloan Digital Sky Survey 
has already been mined to produce a catalog of over 2,500 mostly-new,
mostly-hot white dwarfs \citet{kleinman04}.  These come with  
photometric observations in five ($u$, $g$, $r$, $i$, $z$) bands, and 
with blue and red spectrophotometry.  A limitation is that the 
short-wavelength cutoff of the spectra near $\lambda$ = 3830 \AA\ 
excludes the highest Balmer lines for the bulk of DA white dwarfs, 
and this compromises primarily the measurement of the gravity. 
In any case, it is hoped that this analysis of the PG sample will 
serve as a suitable ``benchmark'' against which the expected 
improvements from such samples as these can be measured.  

The PG sample will retain one advantage over those from the larger,  
deeper surveys: this is a sample which contains many of the brightest, 
hot white dwarfs known.  The availability of accurate parameters 
should support followup observations with {\it FUSE} and other 
future space missions.  Earlier it was mentioned that {\it GALEX} 
may produce a more rigorous determination of the fraction of hot 
white dwarfs hidden by companions. 

\acknowledgments {} We acknowledge S.~Boudreault, A.~Gianninas,
R.~A.~Saffer, and D.~K.~Sing for the acquisition and data reduction of
spectra used in this analysis.  We thank I. Hubeny for calculating LTE
and NLTE model atmospheres for comparison with our calculations.  JL
acknowledges useful conversations with Richard Green concerning the
completeness of the PG, Lars Bildsten concerning the strategy of this
analysis, and Eric Mamajek concerning the Gould's Belt population.  We
thank the referee and Professor Volker Weidemann for careful readings 
of this paper, and for helpful criticisms, suggestions and comments.  
This work was supported in part by the National Science
Foundation through grant AST-0307321, by the NSERC Canada, by the Fund
FQRNT (Qu\'ebec), and by NASA grant NAG5-9408.

\clearpage
\appendix

\section{MASSIVE, EUV-SELECTED WHITE DWARFS AND GOULD'S BELT?} 

We have shown in this paper that, for the PG high Galactic latitude
sample, $\ge80$\% of the DAs with $>0.8$~\msun\ may require a formation
mechanism from binary evolution, rather than from massive (4-8~\msun)
single progenitors.  It is interesting to examine briefly the white
dwarf counterparts found in the $EUVE$ and $ROSAT$ all-sky surveys,
whose parameters are published in \citet{vennes96}, \citet{marsh97},
\citet{vennes97b}, and \citet{vennes99}.  There are 28 white dwarfs from
these sources with estimated masses $>0.8$~\msun, in fact 19 have
$>1$~\msun.  Note that this is a much higher fraction with $>1$~\msun\
than is the case for the PG sample.  All are hotter than 25,000~K, and
all but three are above 30,000~K.  The latter temperature corresponds to
a cooling age near $10^8$ years for a 1~\msun\ white dwarf.  Thus most
of these stars have total ages of a few $\times\ 10^8$ years or less.

These white dwarfs are plotted (with ``star'' symbols) vs. Galactic
coordinates in Figure~\ref{fg:f18}.  While these stars were detected
in an all-sky survey, it is interesting that all but nine are within
$\pm30$ degrees Galactic latitude.  The overlap with PG (whose stars are
all above this latitude) is small.  Most of the massive PG white dwarfs
are too cool to be EUV sources.

The Sun appears to reside currently in a region of enhanced, recent star
formation in the Galactic disk.  Historically, \citet{gould1874} noted
that the brightest stars in the sky lie in a "belt" tilted some
$\sim18$$^o$ from the Galactic plane reaching farthest south near
180$^o$ Galactic longitude.  This distribution, dominated by B and O
stars, is referred to as Gould's Belt.  In Figure~\ref{fg:f18} we
have added as small dots all the B and O stars from the on-line version
of the Yale Bright Star Catalog \citep[]{hoffleit91}.  An analysis of an
earlier version of this sample is presented in \citet{bahcall87}.  The
agreement in positions of the massive EUV white dwarfs and the O,B stars
suggests an association.  The distribution of white dwarfs admittedly
also depends on the very nonuniform interstellar EUV opacity for
distances out to a few hundred parsecs.

A comprehensive analysis of {\it Hipparcos} astrometry by
\citet{dezeeuw99} identified 18 OB associations within this range of
Galactic latitude, located at distances ranging from 100 to roughly 600
pc.  The youngest of these entities such as the Sco~OB2 complex are
probably too young (a few Myr to perhaps 20~Myr) to have produced any
white dwarfs.  However, the older (of order 100~Myr) of these
associations at 100-200~pc distances, such as the $\alpha$~Per and
Cas-Tau associations, may have provided suitable progenitors for massive
white dwarfs.  We thus suggest that a significantly higher fraction of
the massive EUV white dwarfs near the Galactic plane could have evolved
from single-star progenitors than for those at high Galactic latitudes.
Finally, we surmise that the Gould Belt associations may not generally
be old enough for white dwarfs in the 0.8--1.0~\msun\ range to have yet
been produced.

\clearpage
\clearpage
\begin{deluxetable}{llrrlcrclcllc}
\tabletypesize{\scriptsize}
\tablecolumns{12}
\tablewidth{0pt}
\rotate
\tablecaption{Atmospheric Parameters of DA Stars from the Complete PG Sample}
\tablehead{
\colhead{PG} &
\colhead{Name} &
\colhead{$\Te$} &
\colhead{(K)} &
\colhead{log $g$} &
\colhead{$M/$\msun} &
\colhead{$M_V$} &
\colhead{log $L/$\lsun} &
\colhead{$V$} &
\colhead{$D ({\rm pc})$} &
\colhead{$1/V_{\rm max}$} &
\colhead{log $\tau$} &
\colhead{Notes}}
\startdata
0000$+$172&             & 20,210   &(  311)   &7.99 (0.05)    &0.62 (0.03)    &10.70          &$-$1.58        &15.84       &  106&5.77($-$7)     &7.93           &                    \\
0004$+$061&PHL 670      & 15,120   &(  332)   &7.84 (0.05)    &0.52 (0.03)    &10.98          &$-$2.00        &16.23       &  112&9.60($-$7)     &8.26           &1                   \\
0010$+$281&             & 26,280   &(  393)   &7.85 (0.05)    &0.56 (0.02)    & 9.99          &$-$1.03        &15.70       &  139&2.14($-$7)     &7.26           &                    \\
0014$+$098&             & 13,770   &(  342)   &7.96 (0.05)    &0.59 (0.03)    &11.34          &$-$2.24        &15.87       &   80&1.63($-$6)     &8.46           &1, 2                \\
0017$+$061&PHL 790      & 28,060   &(  459)   &7.76 (0.06)    &0.53 (0.03)    & 9.70          &$-$0.85        &14.75       &  102&1.48($-$7)     &7.12           &2                   \\
\\
0033$+$016&G1$-$7       & 10,980   &(  158)   &8.83 (0.05)    &1.12 (0.02)    &13.28          &$-$3.23        &15.61       &   29&2.36($-$5)     &9.32           &                    \\
0048$+$202&             & 20,160   &(  312)   &7.99 (0.05)    &0.62 (0.03)    &10.70          &$-$1.59        &14.85       &   67&5.80($-$7)     &7.93           &                    \\
\enddata
\tablecomments{
(1) Binary; 
(2) $V$ magnitude derived from $B_{\rm ph}$; 
(3) DAO star, $\Te$ and $\logg$ from \citealt{bergeron94}; 
(4) Misclassified in the PG catalog; 
(5) $V$ magnitude derived from $g$ and $r$ Sloan Digital Sky Survey; 
(6) ZZ Ceti star; 
(7) Double degenerate; 
(8) DA + magnetic DB,                                              $\Te$ and $\logg$ from \citealt{liebert93}; 
(9) Planetary nebula nucleus; 
(10) Magnetic, $\Te$ and $\logg$ from \citealt{schmidt86}; 
(11) Magnetic, $\Te$ and $\logg$ from \citealt{bergeron02}; 
(12) Magnetic, $\Te$ and $\logg$ from \citealt{liebert03}; 
(13) Double degenerate?; 
(14) Magnetic; 
(15) Magnetic, $\Te$ and $\logg$ from \citealt{liebert85}; 
(16) Magnetic, $\Te$ and $\logg$ from \citealt{schmidt92}; 
(17) Not in \citealt{mccook99}. }
\end{deluxetable}


\clearpage
\begin{deluxetable}{llrrlcrclcllc}
\tabletypesize{\scriptsize}
\tablecolumns{9}
\tablewidth{0pt}
\tablecaption{Atmospheric Parameters of other DA  Stars from the PG Survey}
\tablehead{
\colhead{PG} &
\colhead{Name} &
\colhead{$\Te$} &
\colhead{(K)} &
\colhead{log $g$} &
\colhead{$M/$\msun} &
\colhead{$M_V$} &
\colhead{log $L/$\lsun} &
\colhead{log $\tau$} &
\colhead{Notes}}
\startdata
0019$+$150&PHL 802      & 30,800   &(  475)   &8.10 (0.06)    &0.71 (0.03)    &10.03          &$-$0.90        &7.14           &                    \\
0101$+$059&PB 6250      & 14,190   &(  394)   &8.35 (0.05)    &0.83 (0.03)    &11.86          &$-$2.43        &8.66           &                    \\
0232$+$035&Feige 24     & 65,290   &( 1553)   &7.24 (0.08)    &0.47 (0.02)    & 7.39          &$+$1.09        &5.40           &                    \\
0302$+$028&GD 41        & 35,230   &(  528)   &7.78 (0.05)    &0.55 (0.02)    & 9.24          &$-$0.45        &6.78           &1                   \\
0307$+$149&             & 21,320   &(  347)   &8.03 (0.05)    &0.64 (0.03)    &10.66          &$-$1.51        &7.87           &                    \\
\\
0308$+$096&             & 25,900   &(  451)   &8.08 (0.06)    &0.68 (0.03)    &10.37          &$-$1.20        &7.54           &                    \\
0308$+$188&             & 17,860   &(  264)   &7.85 (0.04)    &0.54 (0.02)    &10.71          &$-$1.72        &8.01           &                    \\
\enddata
\tablecomments{
(1) Classified sd in the PG catalog; 
(2) Classified DB in the PG catalog. }
\end{deluxetable}


\clearpage
\begin{deluxetable}{lllll}

\tablenum{3}
\tablecaption{White Dwarf Evolution Rates} 
\tablewidth{0pt}
\tablehead{
\colhead{log~$\tau$} & \colhead{Delta~$\tau$} & \colhead{Main} & \colhead{Low} &
\colhead{High} }

\startdata

6.25 & $2.16\,(6)$ & $4.21\,(-13)$ & $0.00     $ & $4.83\,(-14)$ \\
6.75 & $6.84\,(6)$ & $4.43\,(-13)$ & $2.75\,(-14)$ & $0.00     $ \\
7.25 & $2.16\,(7)$ & $5.06\,(-13)$ & $4.72\,(-14)$ & $0.00     $ \\
7.75 & $6.84\,(7)$ & $3.95\,(-13)$ & $1.66\,(-14)$ & $7.79\,(-14)$ \\
8.25 & $2.16\,(8)$ & $3.28\,(-13)$ & $1.01\,(-14)$ & $7.67\,(-14)$ \\
8.75 & $6.84\,(8)$ & $1.72\,(-13)$ & $3.31\,(-15)$ & $3.72\,(-14)$ \\
9.25 & $2.16\,(9)$ & $1.45\,(-13)$ & $0.00     $ & $1.41\,(-13)$ \\

\enddata
\end{deluxetable}


\clearpage

\figcaption[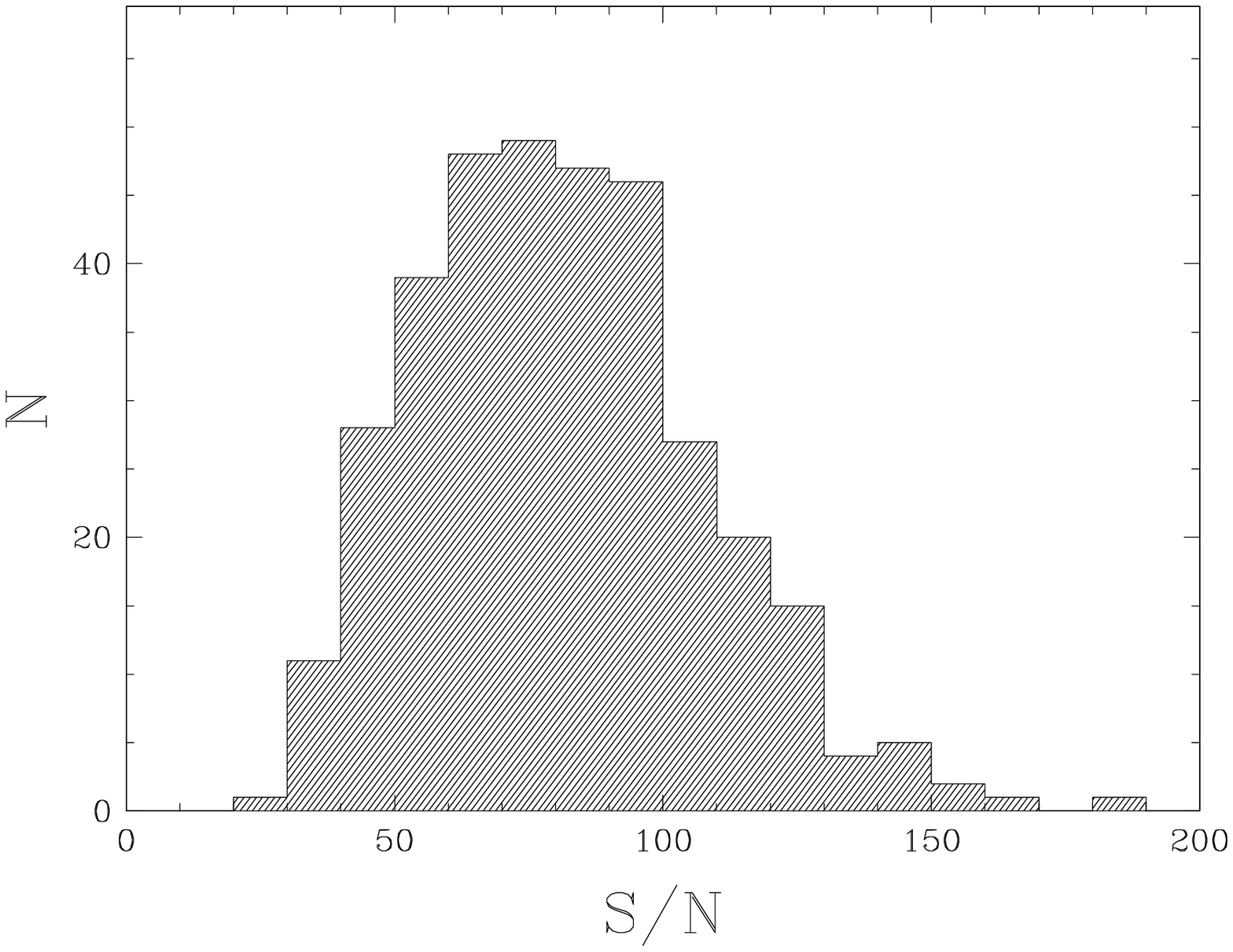] {Distribution of the signal-to-noise ratio for the 
345 optical spectra secured for the analysis of the complete PG
sample.\label{fg:sn}}

\figcaption[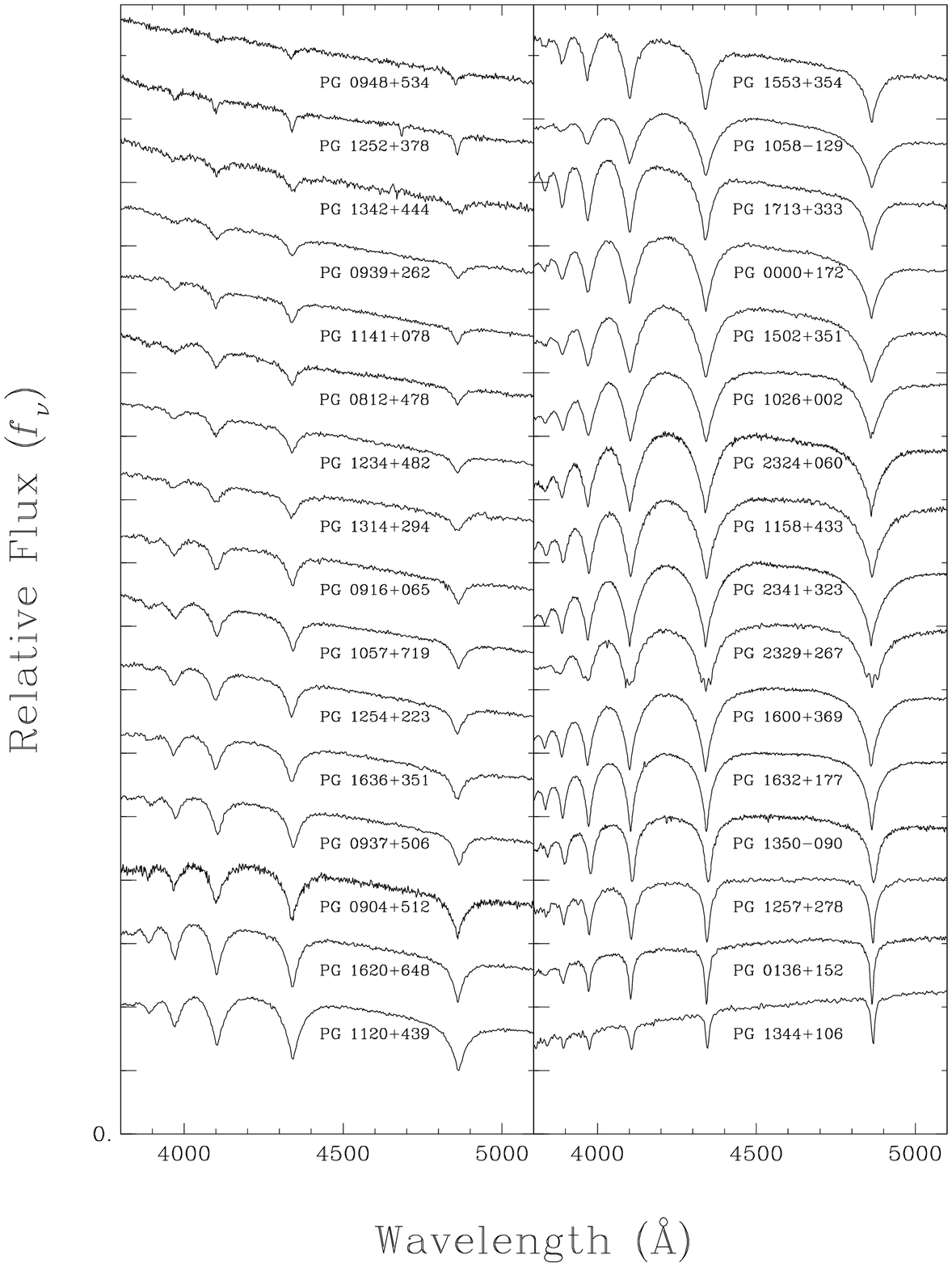] {Optical spectra for a subsample of
DA stars from the PG catalog. The spectra are normalized at 4500~\AA,
and are shifted vertically; the various zero points are indicated by
long tick marks. The effective temperature decreases from upper left
to bottom right.\label{fg:sample}}

\figcaption[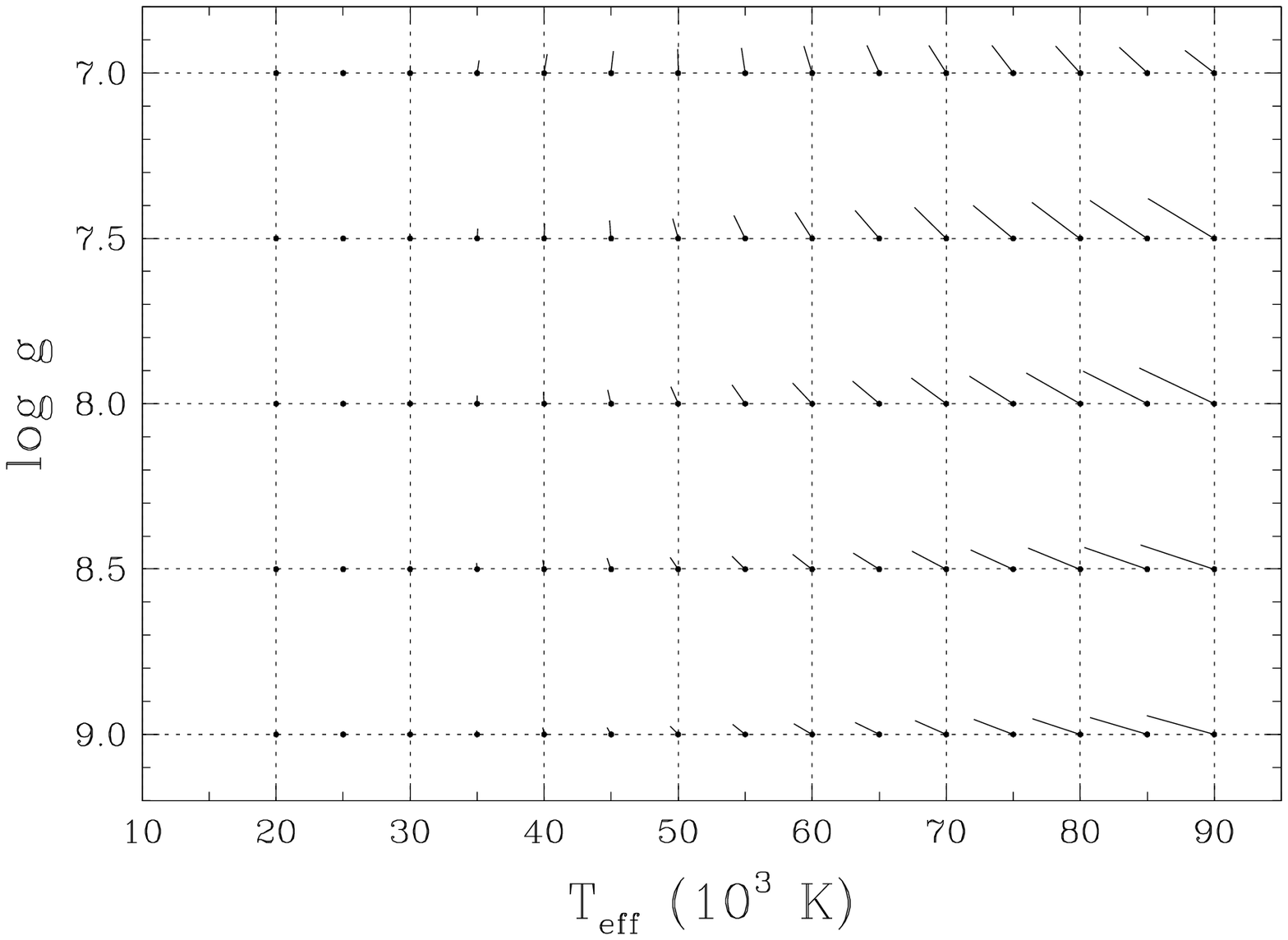] {Corrections that must be applied to transform 
atmospheric parameters obtained from LTE model spectra to account for NLTE
effects. The differences are magnified three times in the $\logg$
direction. LTE models tend to overestimate both $\Te$ and
$\logg$.\label{fg:NLTEcorr}}

\figcaption[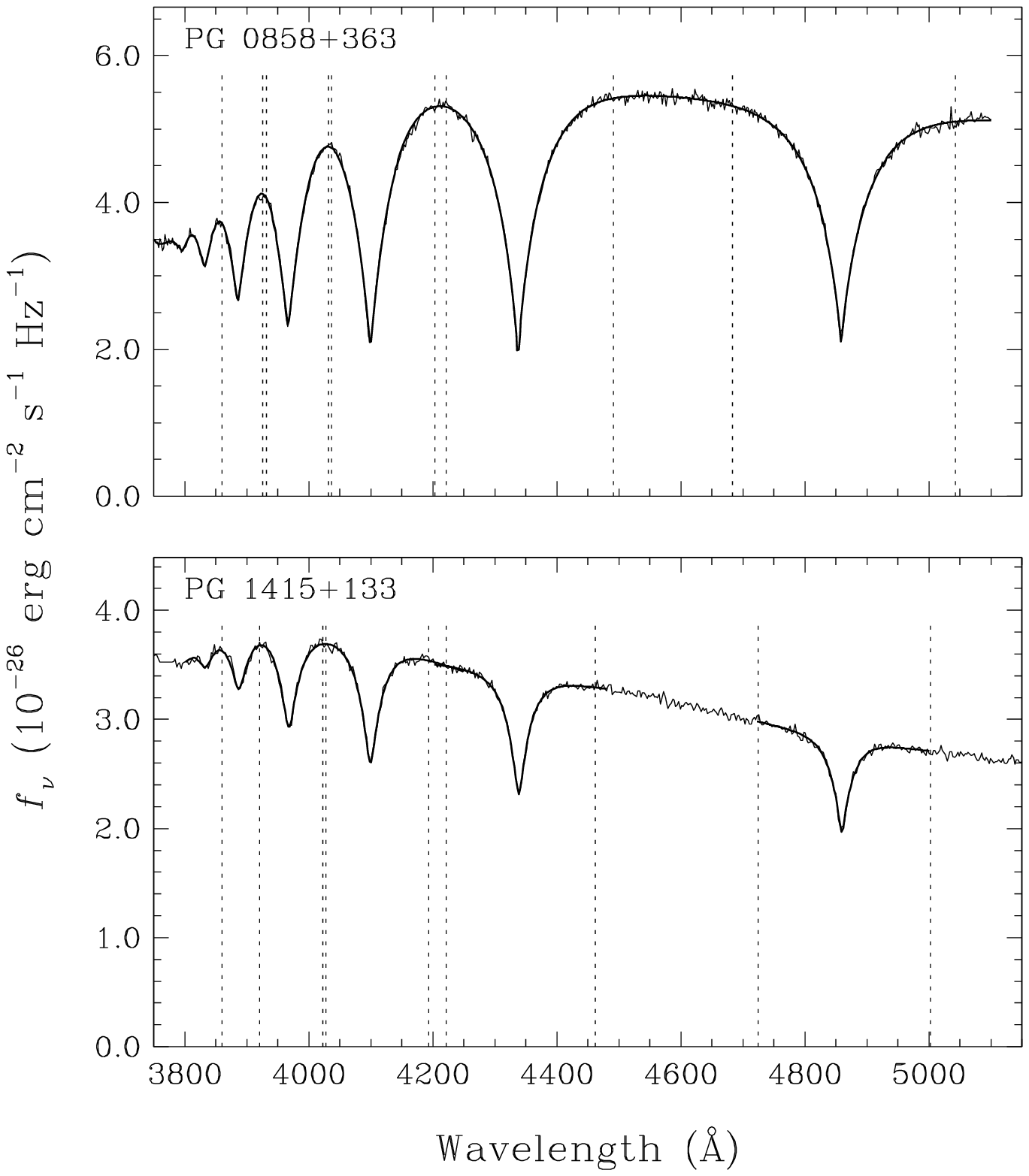] {Examples of the procedures used to define the 
continuum with pseudo-Gaussian profiles ({\it top panel};
$\Te\sim12,000$~K) or model spectra ({\it bottom panel};
$\Te\sim35,000$~K). See text for details. The dotted lines indicate
the wavelength range used to define the continuum for each line.
\label{fg:f4}}

\figcaption[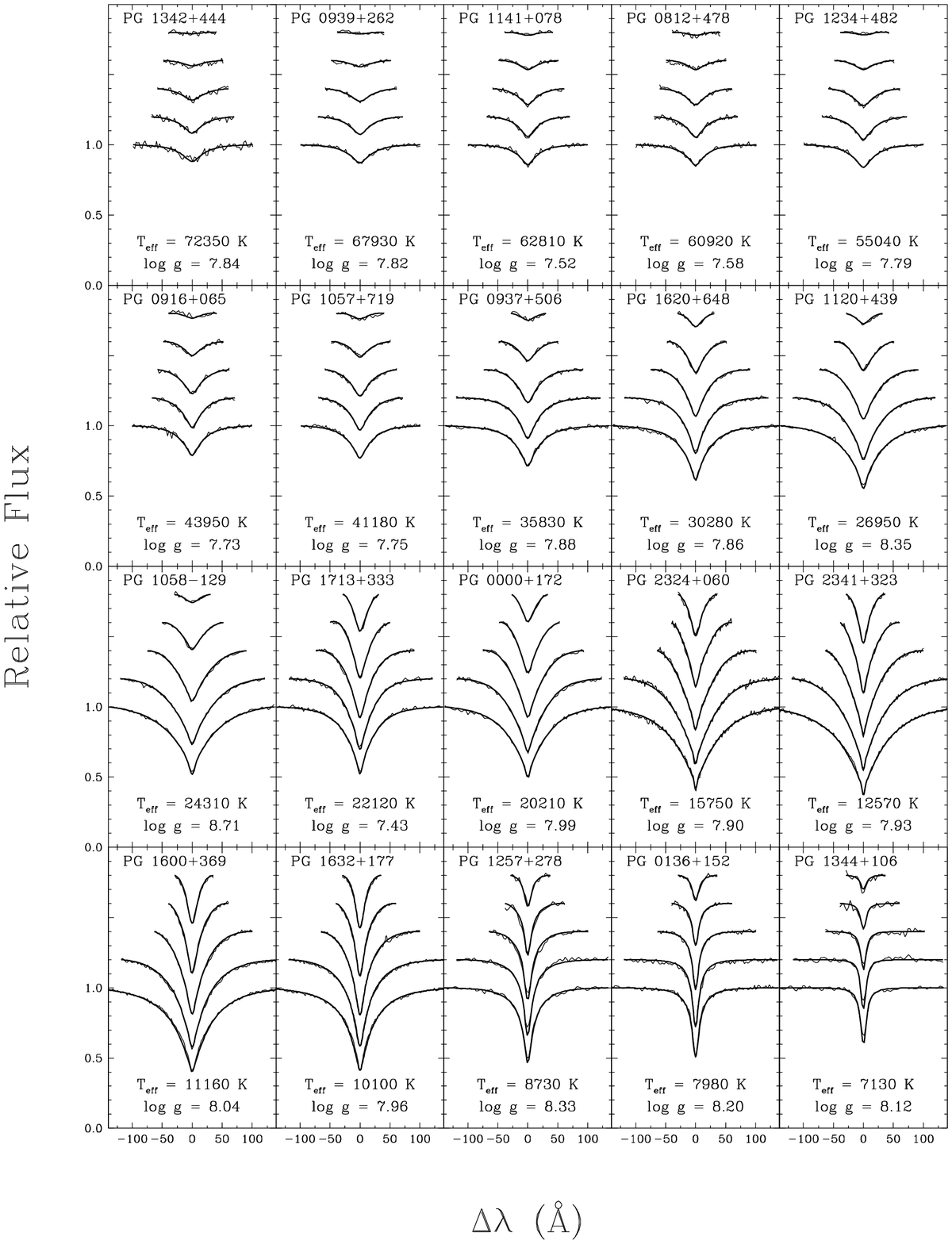] {Model fits to the individual Balmer line profiles
of sample DA stars taken from Figure \ref{fg:sample} in order of
decreasing effective temperature. The lines range from H$\beta$ ({\it
bottom}) to H8 ({\it top}), each offset vertically by a factor of
0.2.\label{fg:f5}}

\figcaption[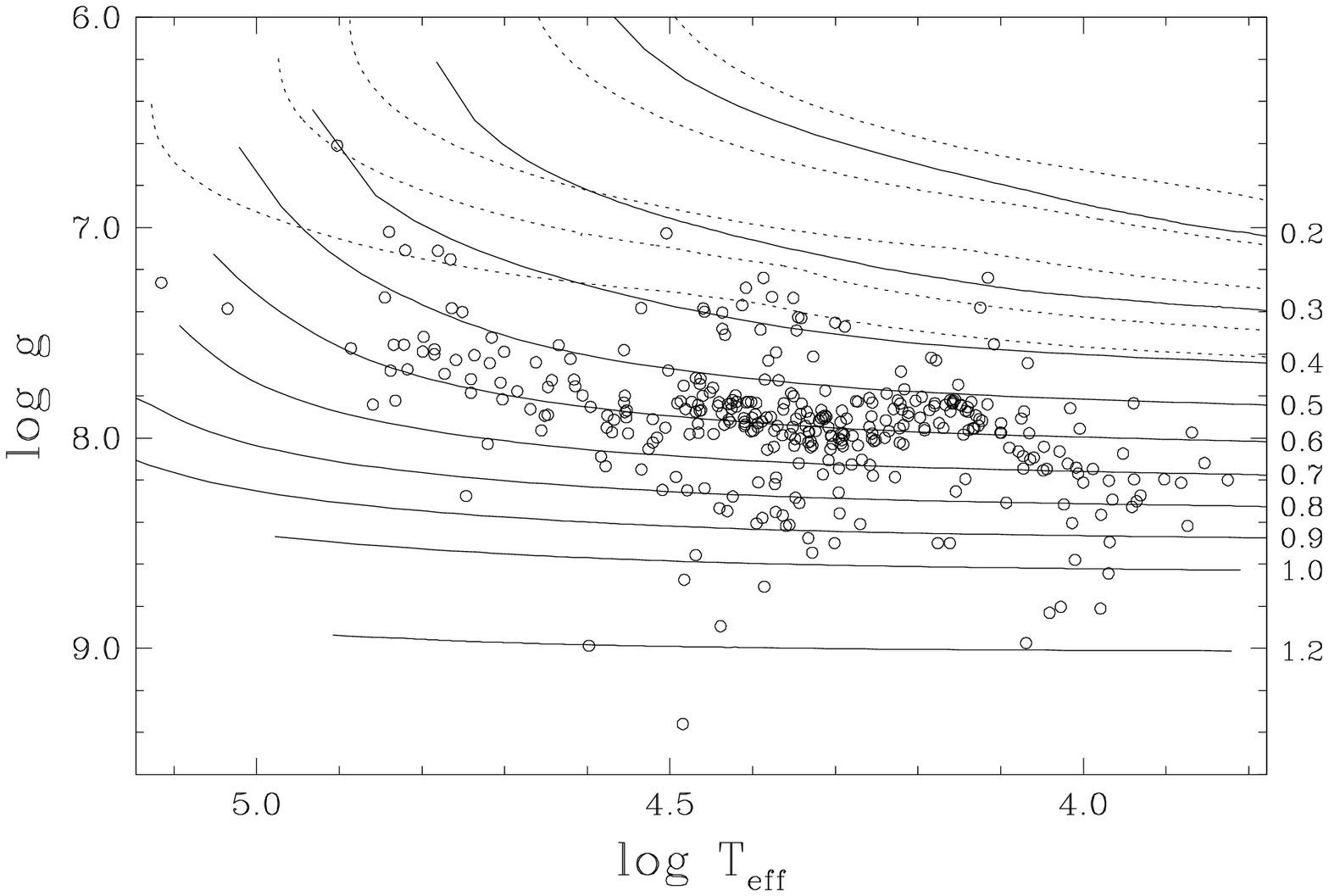] {$\Te$ and $\logg$ values for all DA stars 
from the complete PG sample. The solid lines represent the carbon-core
evolutionary models of \citet{wood95} with thick hydrogen layers;
numbers on the right hand side of the figure indicate the mass of each
model in solar masses. The dotted lines represent the helium-core
models of \citet{althaus01} for, from top to bottom, $M=0.196$, 0.242, 
0.292, 0.360, and 0.406 \msun.\label{fg:f6}}

\figcaption[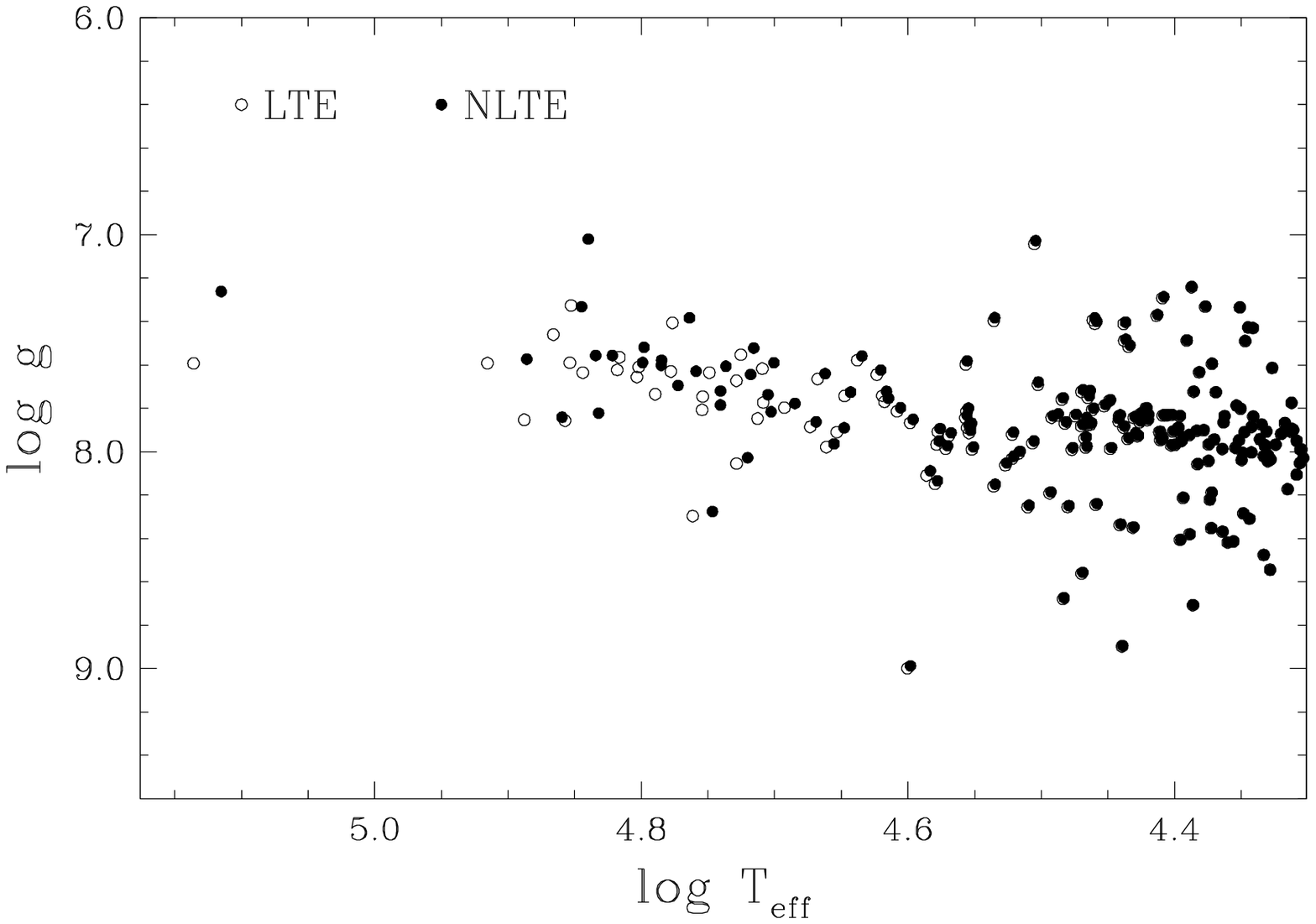] {Comparison of atmospheric parameter solutions 
obtained under the assumption of LTE ({\it open circles}) and NLTE
({\it filled circles}).\label{fg:f7}}

\figcaption[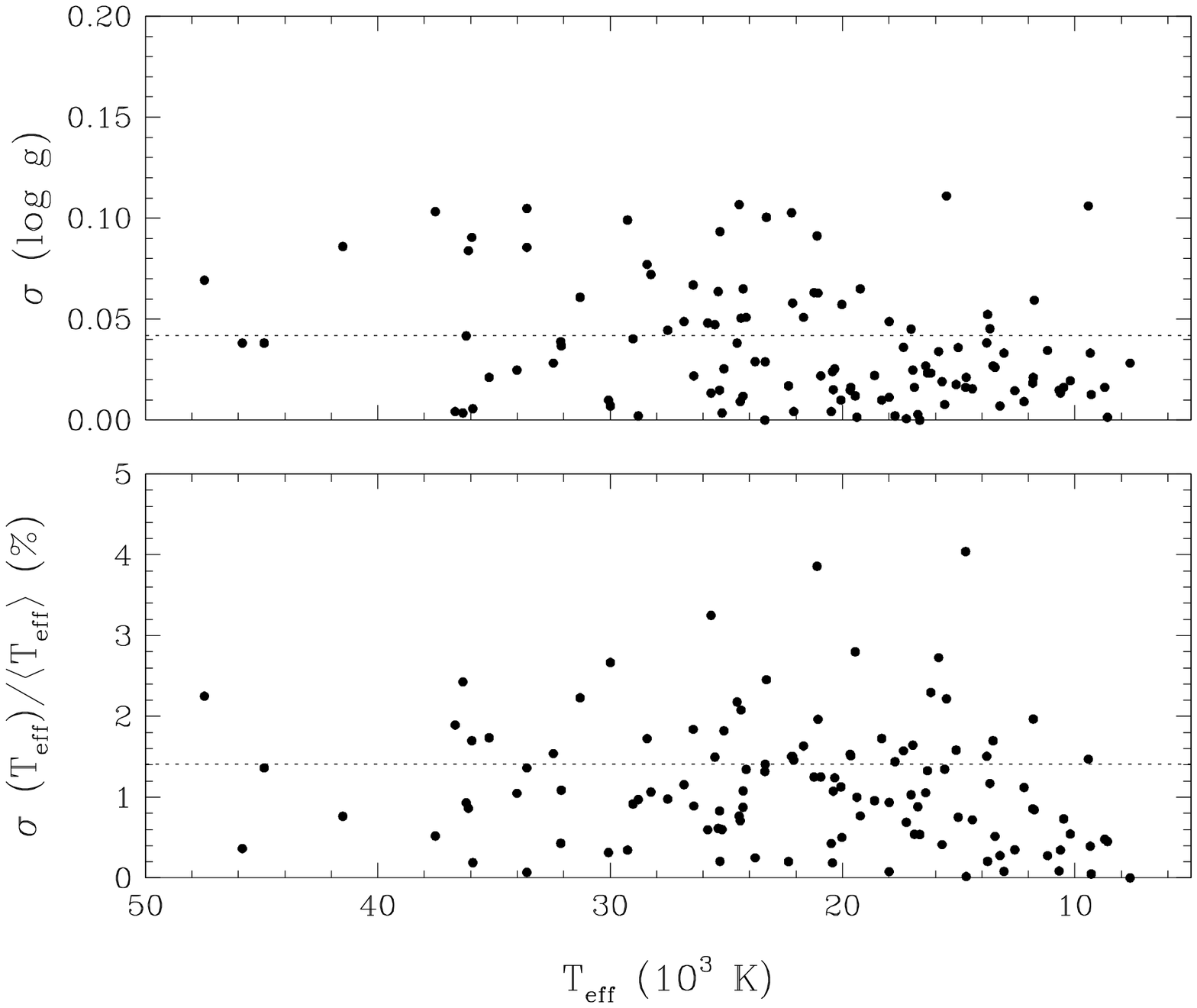] {Distribution of standard deviations 
in $\Te$ and $\logg$ for individual white dwarf stars with multiple
measurements, as a function of effective temperature. Standard
deviations in $\Te$ are expressed in percentage with respect to the
average temperature of the star. The dotted lines represent the
average standard deviations.\label{fg:multiple}}

\figcaption[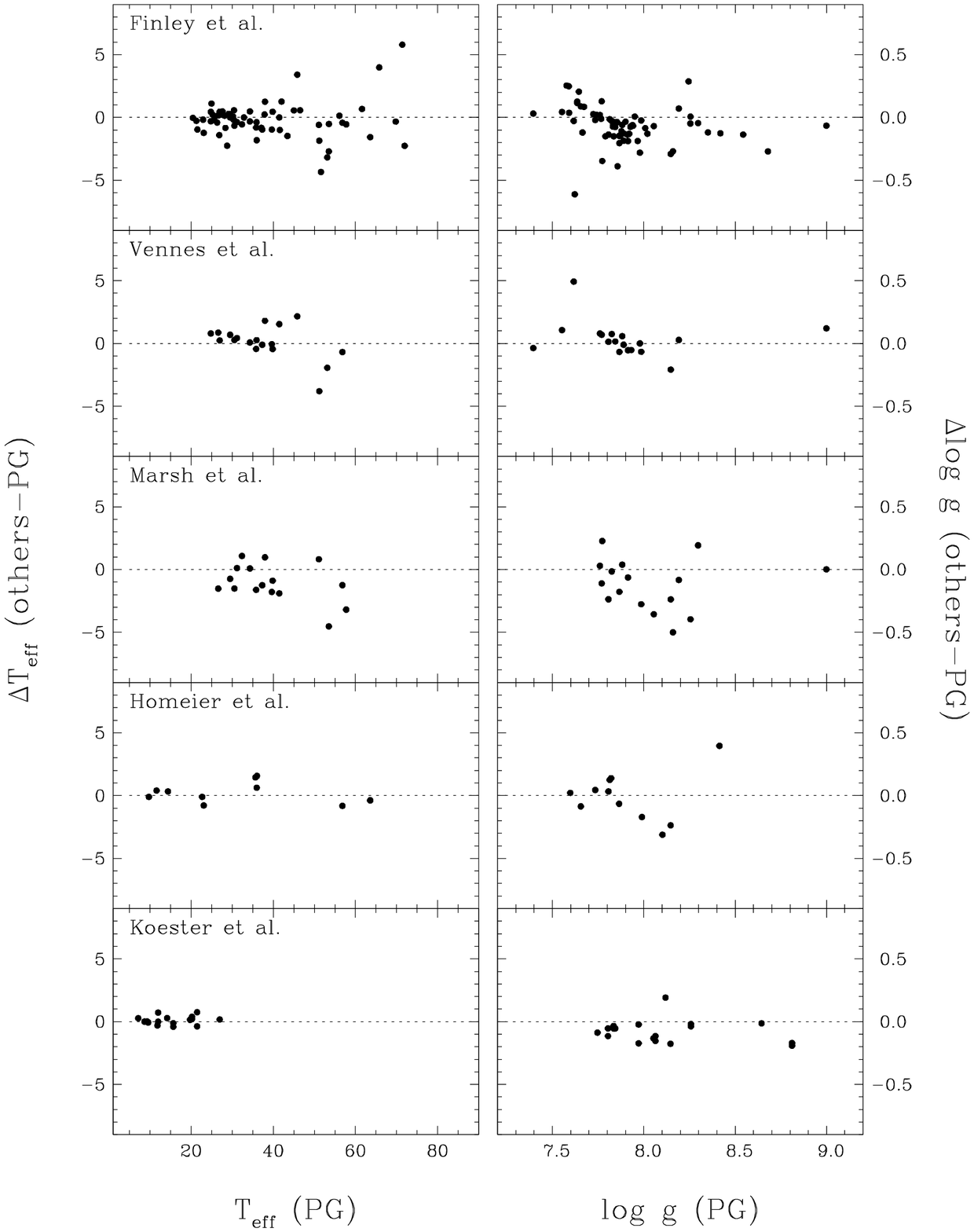] {Comparison of our $\Te$ and $\logg$
determinations of PG stars in common with the data sets of
\citet{finley97}, \citet{vennes97b}, \citet{marsh97}, \citet{homeier98},
and \citet{koester01}. In each panel we plot the differences between
these investigations ({\it others}) and our results (PG) as a
function of our determinations of $\Te$ or $\logg$. Effective
temperatures are in units of $10^3$ K. \label{fg:f9}}

\figcaption[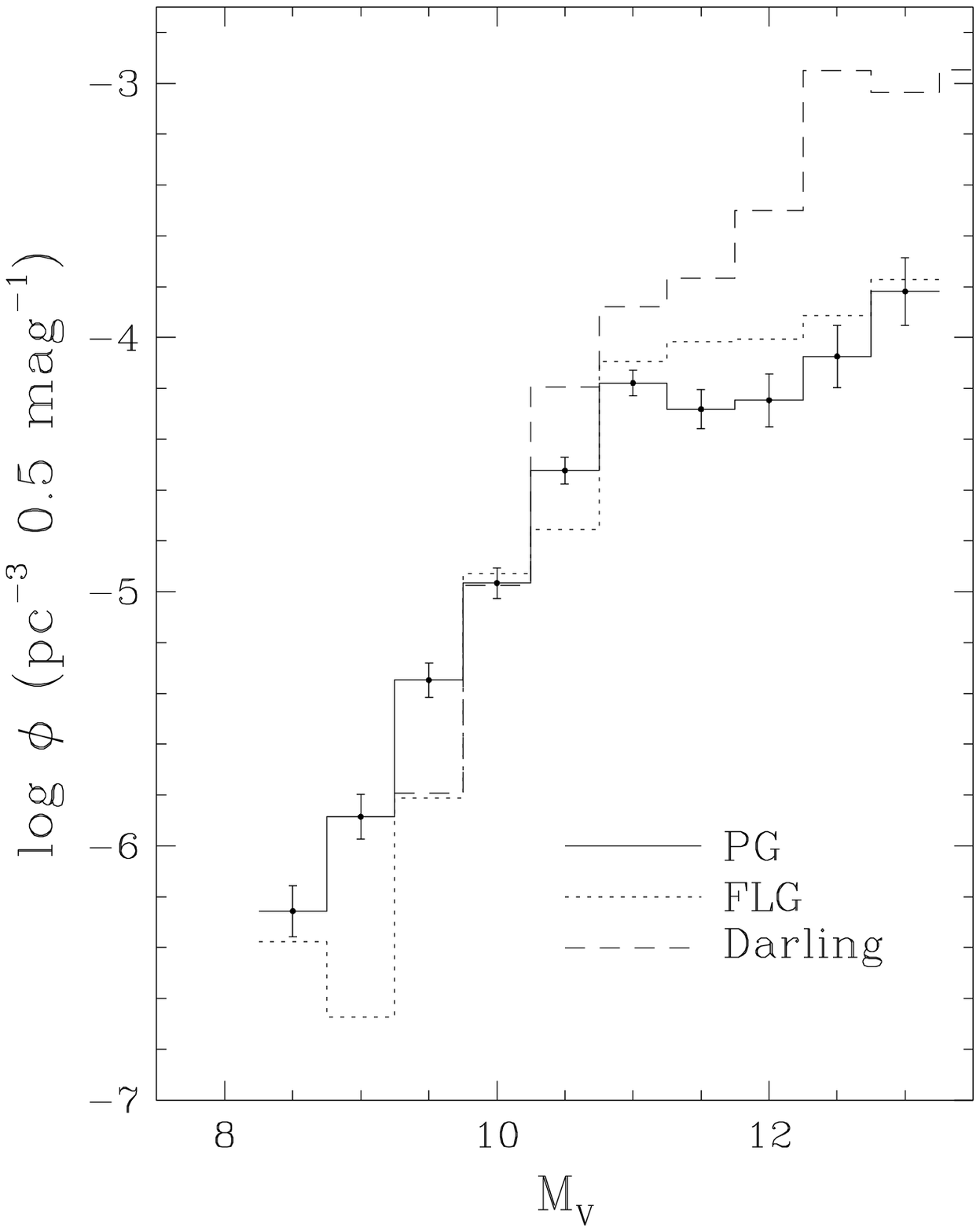] {Luminosity function of all DA stars in the
complete PG sample ({\it solid line}) presented in half-magnitude bins,
assuming a scale height for the Galaxy of $z_0=250$ pc. The dotted line
represents the results of \citet{fleming86}, and the dashed line
that of \citet{darling94}, shown here for comparison.\label{fg:f10}}

\figcaption[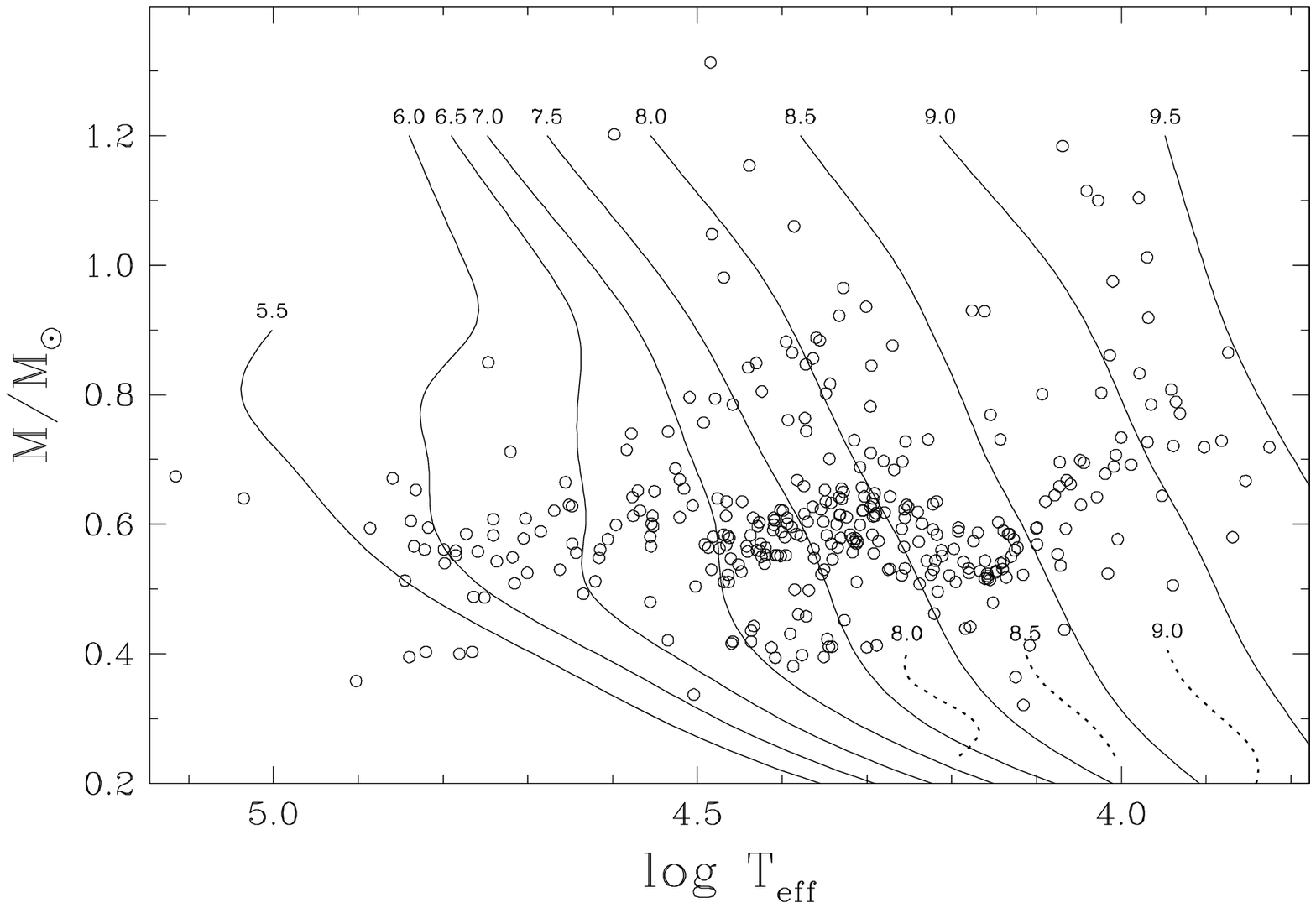] {Masses of all DA stars in the complete PG sample
as a function of $\log\,\Te$, together with the theoretical isochrones
from \citet[][ {\it solid lines}]{wood95} and \citet[][ {\it dotted
lines}]{althaus01}. Isochrones are labeled in units of log $\tau$,
where $\tau$ is the white dwarf cooling age in years. \label{fg:f11}}

\figcaption[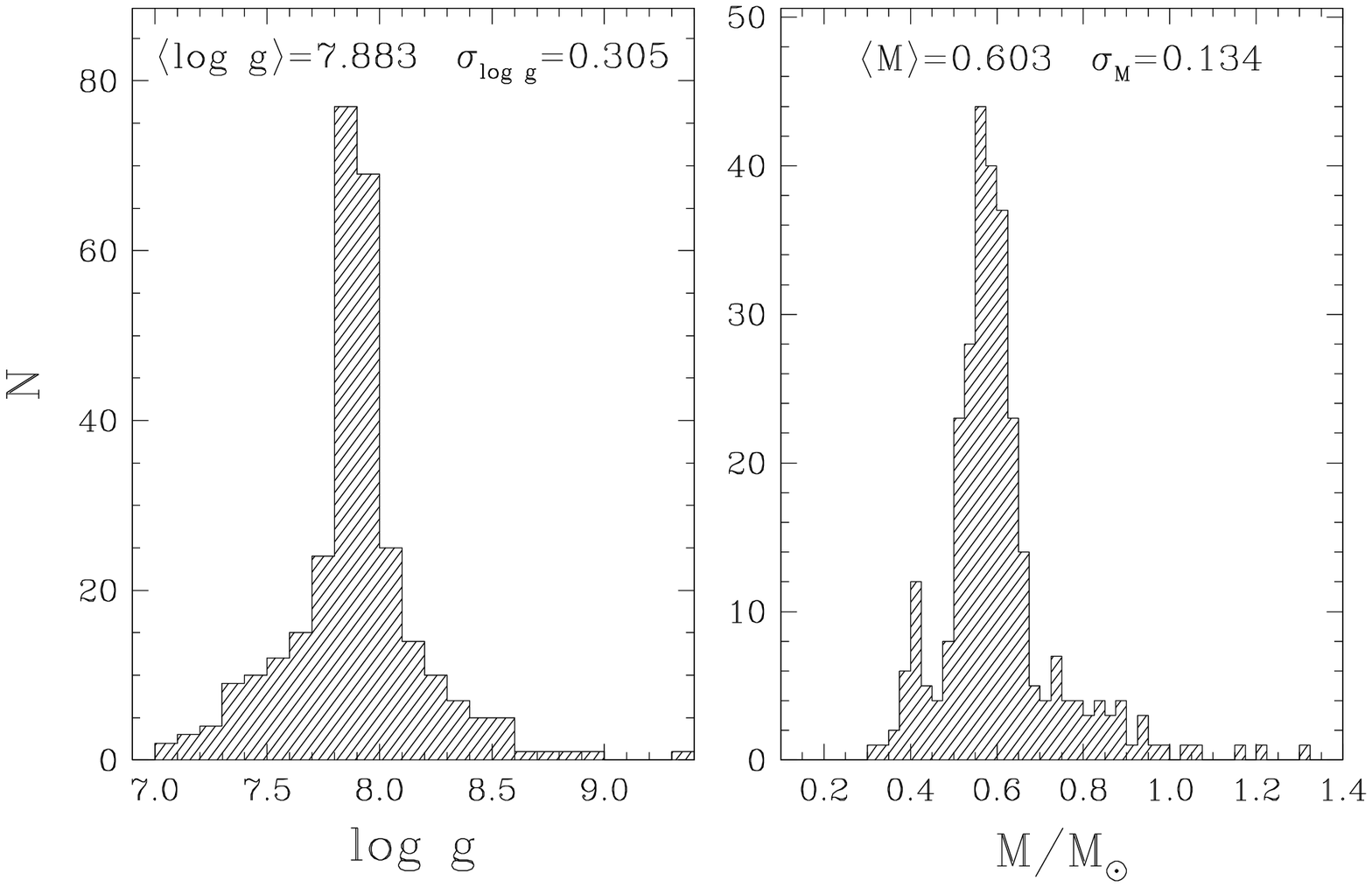] {Surface gravity and mass distributions for the 297
DA stars in the complete PG sample with $\Te\ >$13,000~K. The masses of
DA stars below this value may be biased, as explained in the text.  Mean
values and standard deviations of both distributions are indicated in
each panel.
\label{fg:f12}}

\figcaption[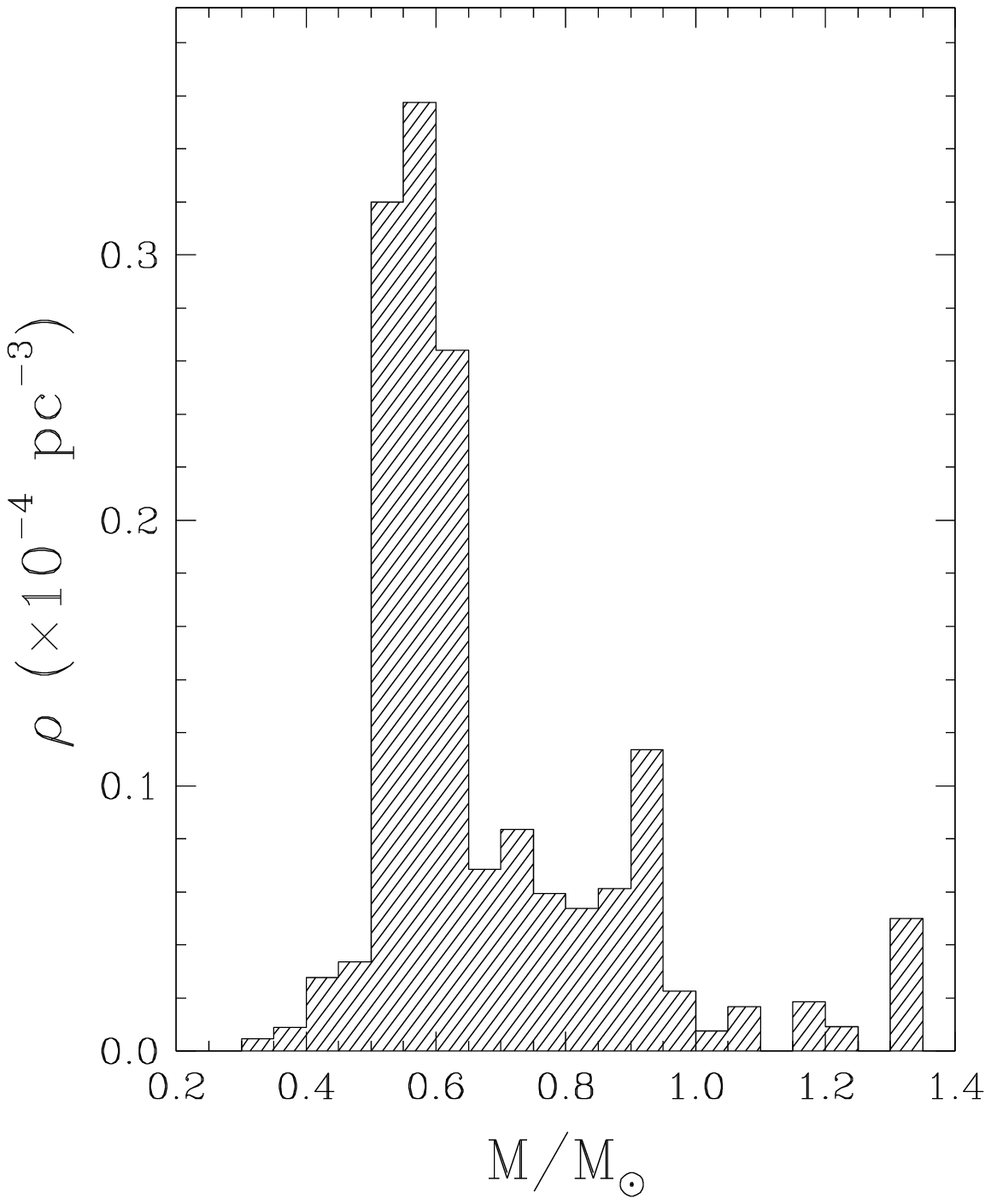] {The $1/V_{\rm max}$ weighted mass distribution 
of the 297 PG DA stars with $\Te\ >$13,000~K.
\label{fg:f13}} 

\figcaption[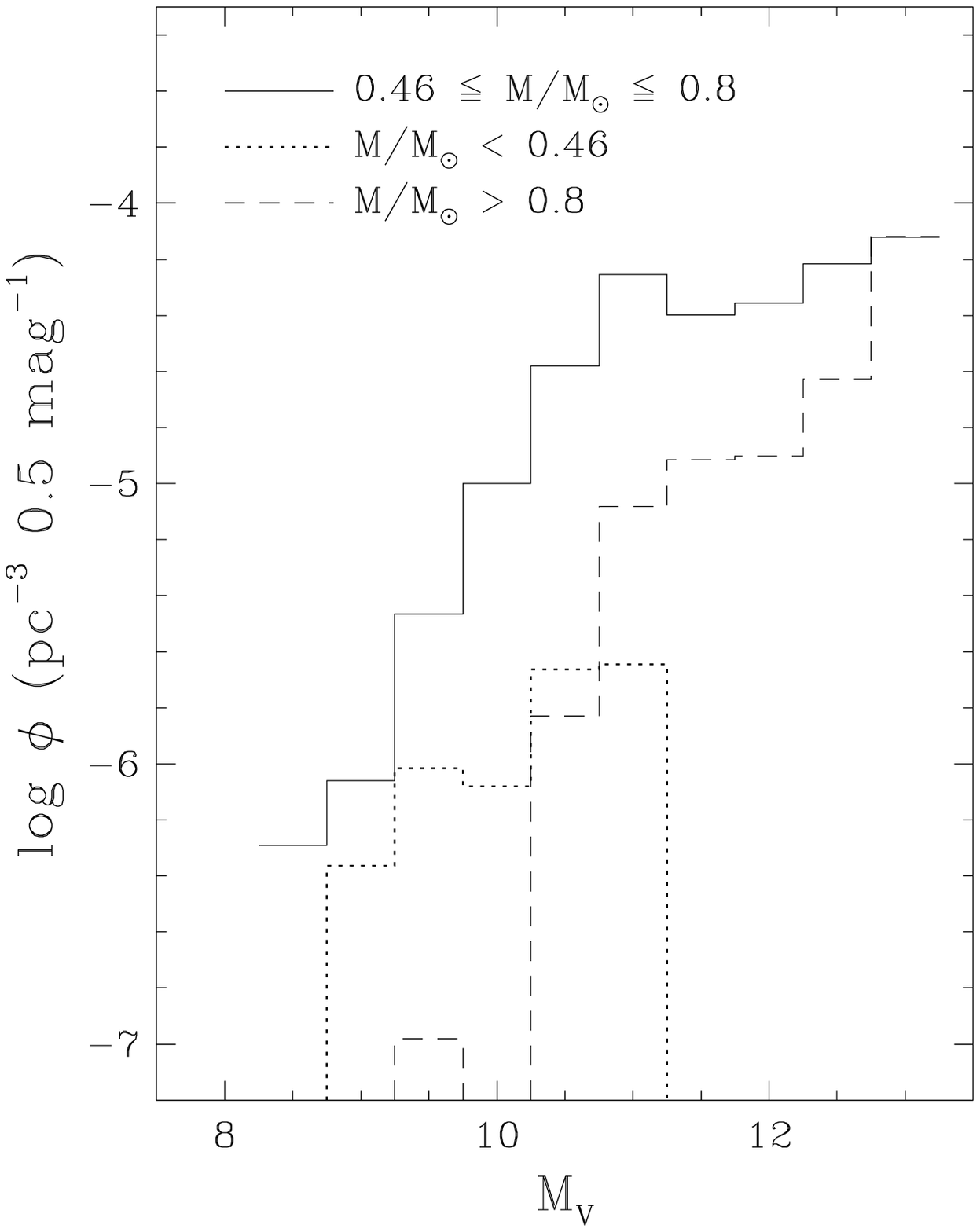]{The luminosity functions of the three 
arbitrary mass components as discussed in the text, with symbols as 
shown in the figure. \label{fg:f14}}

\figcaption[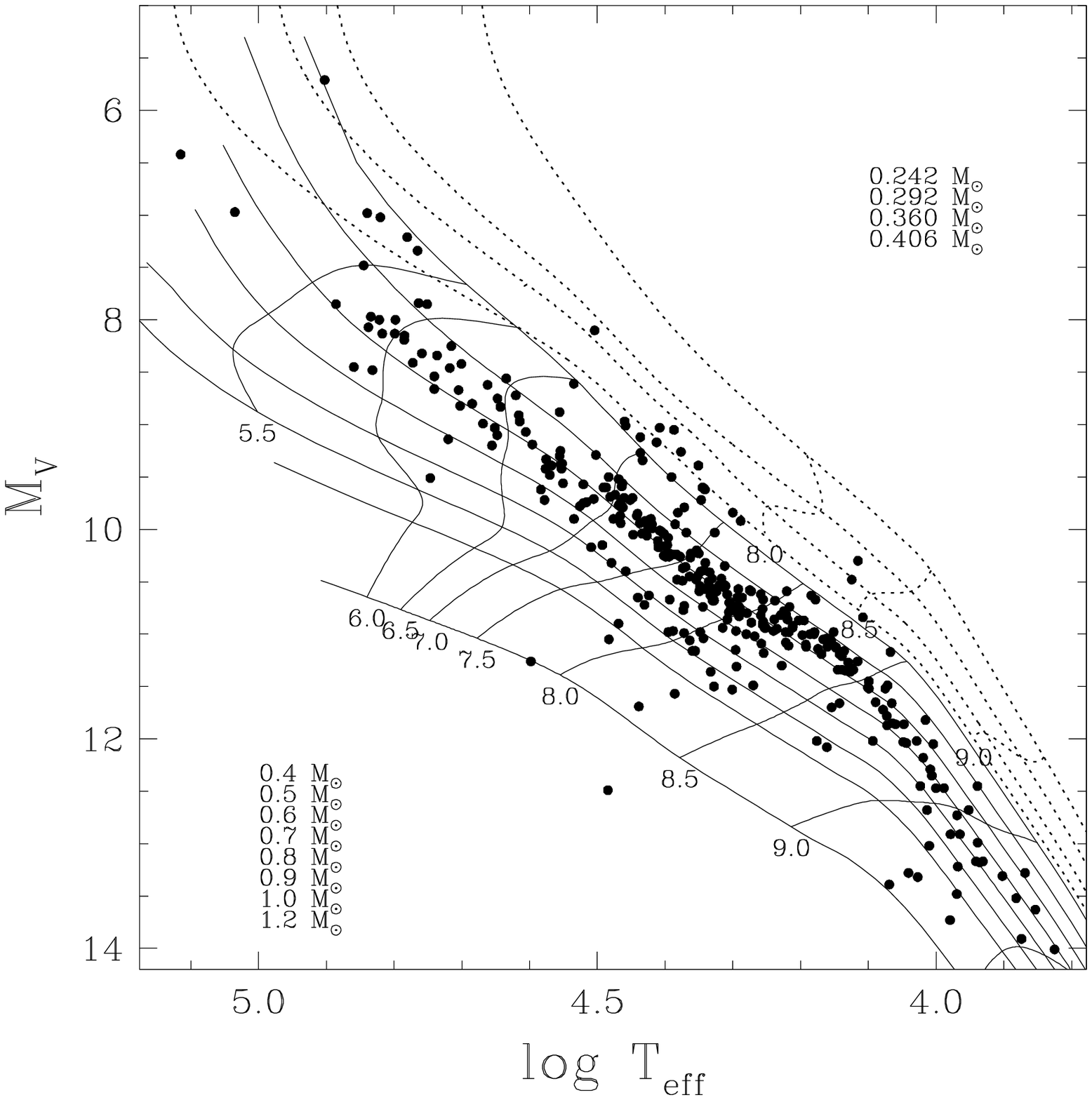] {Absolute visual magnitudes $\mv$ as a function 
of $\log\,\Te$ for all DA stars in the complete PG sample. The
theoretical isochrones from \citet[][ {\it solid lines}]{wood95} and
\citet[][ {\it dotted lines}]{althaus01} are also shown, with corresponding 
masses indicated in the Figure. Isochrones are labeled in units of log
$\tau$, where $\tau$ is the white dwarf cooling age in years.
\label{fg:f15}}

\figcaption[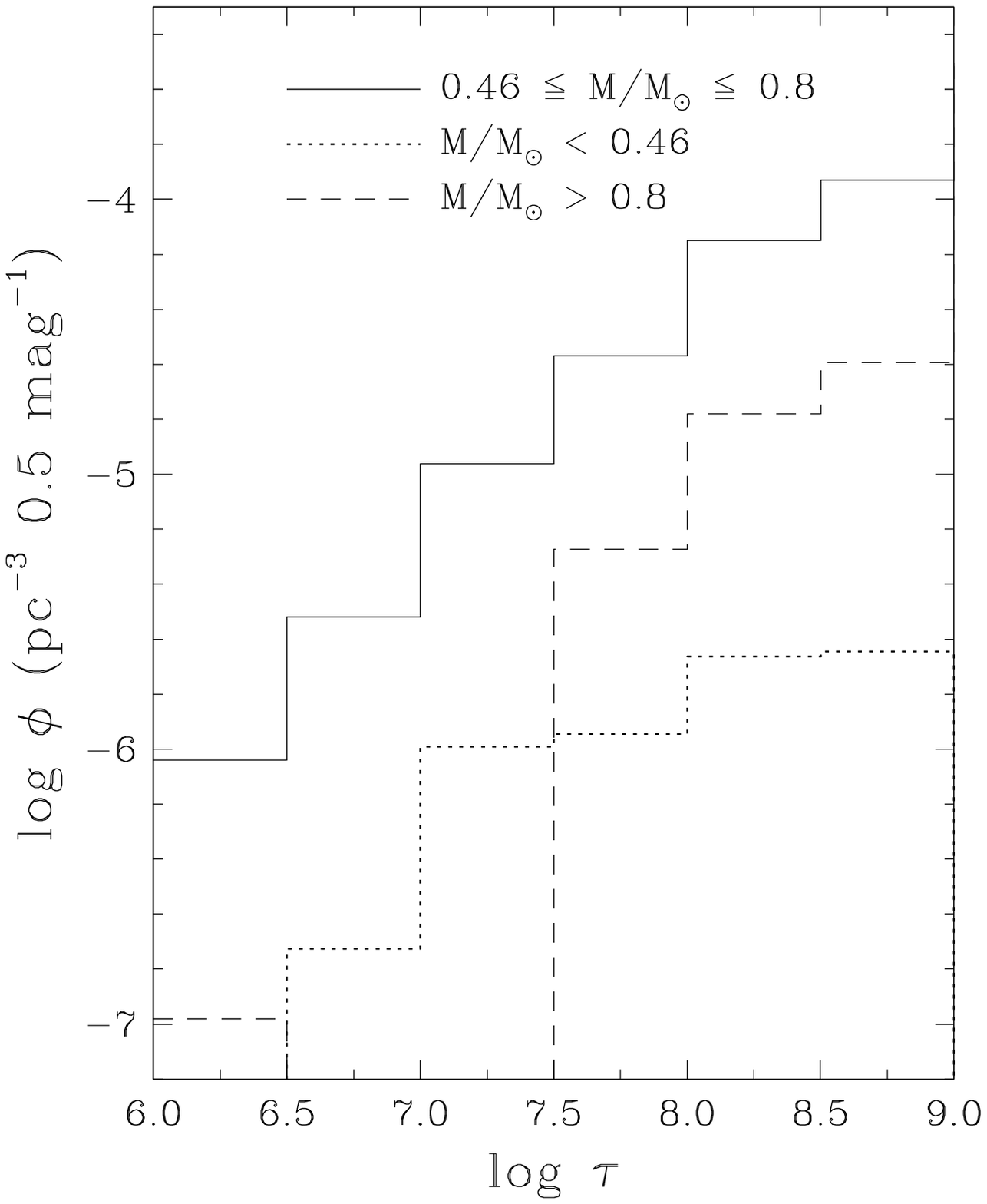] {The number density of white dwarf weighted by 
$1/V_{\rm max}$ is plotted for 0.5 dex intervals of the log of the cooling 
time, for the three mass components shown in \label{fg:f16}}

\figcaption[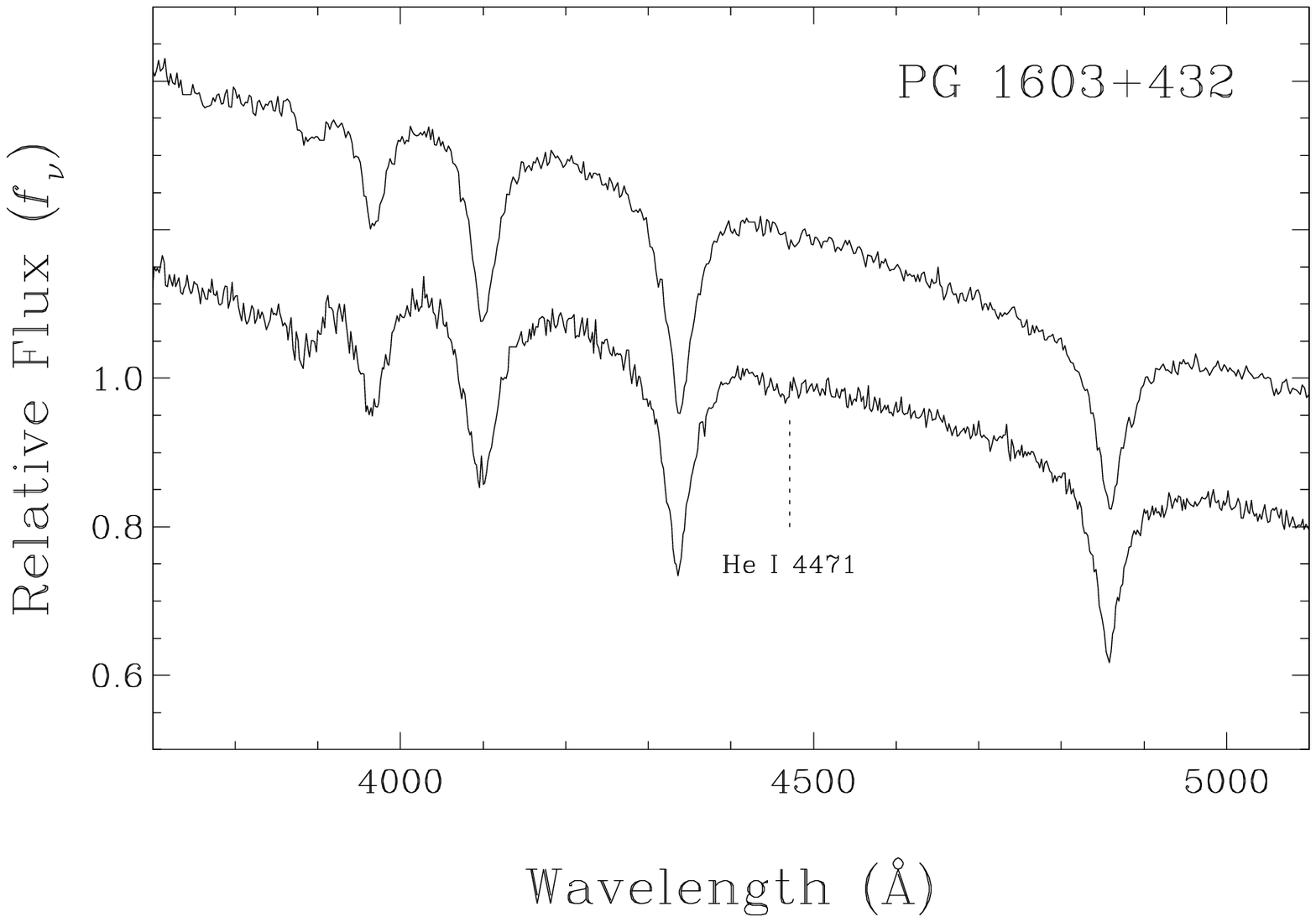] {The spectrum published in \citet{bergeron94}
({\it bottom}) is displayed with a new, higher signal-to-noise ratio spectrum
({\it top}). Both appear to show a weak line due to He~I 4471~\AA.  No other 
He~I or He~II features appear. \label{fg:f17}}

\figcaption[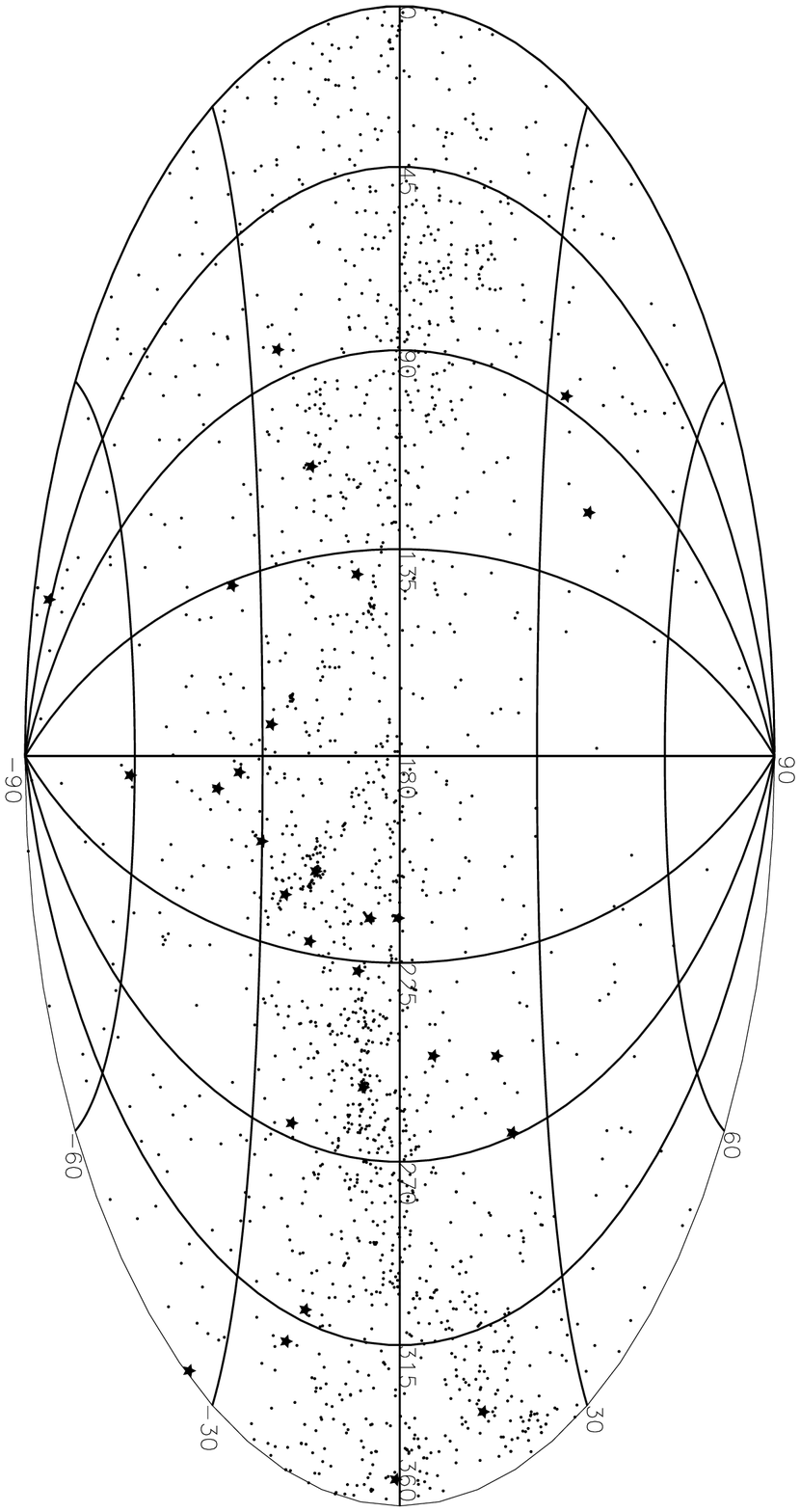] {The 27 white dwarfs found in the $EUVE$ and
$ROSAT$ all-sky surveys with mass estimates $>$0.8~\msun\ are shown as
``star'' symbols, plotted in Galactic coordinates.  Also shown as small
dots are the positions of all O and B stars in the on-line Yale Bright
Star Catalog.  \label{fg:f18}}

\clearpage
\begin{figure}[p]
\plotone{f1.eps}
\begin{flushright}
Figure \ref{fg:sn}
\end{flushright}
\end{figure}

\clearpage
\begin{figure}[p]
\plotone{f2.eps}
\begin{flushright}
Figure \ref{fg:sample}
\end{flushright}
\end{figure}

\clearpage
\begin{figure}[p]
\plotone{f3.eps}
\begin{flushright}
Figure \ref{fg:NLTEcorr}
\end{flushright}
\end{figure}

\clearpage
\begin{figure}[p]
\plotone{f4.eps}
\begin{flushright}
Figure \ref{fg:f4}
\end{flushright}
\end{figure}

\clearpage
\begin{figure}[p]
\plotone{f5.eps}
\begin{flushright}
Figure \ref{fg:f5}
\end{flushright}
\end{figure}

\clearpage
\begin{figure}[p]
\plotone{f6.eps}
\begin{flushright}
Figure \ref{fg:f6}
\end{flushright}
\end{figure}

\clearpage
\begin{figure}[p]
\plotone{f7.eps}
\begin{flushright}
Figure \ref{fg:f7}
\end{flushright}
\end{figure}

\clearpage
\begin{figure}[p]
\plotone{f8.eps}
\begin{flushright}
Figure \ref{fg:multiple}
\end{flushright}
\end{figure}

\clearpage
\begin{figure}[p]
\plotone{f9.eps}
\begin{flushright}
Figure \ref{fg:f9}
\end{flushright}
\end{figure}

\clearpage
\begin{figure}[p]
\plotone{f10.eps}
\begin{flushright}
Figure \ref{fg:f10}
\end{flushright}
\end{figure}

\clearpage
\begin{figure}[p]
\plotone{f11.eps}
\begin{flushright}
Figure \ref{fg:f11}
\end{flushright}
\end{figure}

\clearpage
\begin{figure}[p]
\plotone{f12.eps}
\begin{flushright}
Figure \ref{fg:f12}
\end{flushright}
\end{figure}

\clearpage
\begin{figure}[p]
\plotone{f13.eps}
\begin{flushright}
Figure \ref{fg:f13}
\end{flushright}
\end{figure}

\clearpage
\begin{figure}[p]
\plotone{f14.eps}
\begin{flushright}
Figure \ref{fg:f14}
\end{flushright}
\end{figure}

\clearpage
\begin{figure}[p]
\plotone{f15.eps}
\begin{flushright}
Figure \ref{fg:f15}
\end{flushright}
\end{figure}

\clearpage
\begin{figure}[p]
\plotone{f16.eps}
\begin{flushright}
Figure \ref{fg:f16}
\end{flushright}
\end{figure}

\clearpage
\begin{figure}[p]
\plotone{f17.eps}
\begin{flushright}
Figure \ref{fg:f17}
\end{flushright}
\end{figure}

\clearpage
\begin{figure}[p]
\plotone{f18.eps}
\begin{flushright} 
Figure \ref{fg:f18} 
\end{flushright}
\end{figure} 

\end{document}